\documentclass[%
 reprint,
 amsmath,amssymb,
 aps,
]{revtex4-2}

\usepackage{graphicx}
\usepackage{dcolumn}
\usepackage{bm}
\usepackage{xcolor}
\usepackage{physics}
\usepackage{hyperref}
\usepackage{cleveref}

\newcommand{\titleplaceholder}{Engineering entanglement geometry via spacetime-modulated measurements}

\newcommand{\diff}{\ensuremath{{\rm d}}} 

\begin{document}

\preprint{APS/123-QED}

\title{\titleplaceholder} 

\author{Aditya Cowsik}
\email{acowsik@stanford.edu}
\affiliation{Department of Physics, Stanford Unviersity, Stanford, CA, 94305}
\author{Matteo Ippoliti}
\affiliation{Department of Physics, Stanford Unviersity, Stanford, CA, 94305}
\author{Xiao-Liang Qi}
\affiliation{Department of Physics, Stanford Unviersity, Stanford, CA, 94305}

\date{\today}

\begin{abstract}
We introduce a general approach to realize quantum states with holographic entanglement structure via monitored dynamics. Starting from random unitary circuits in $1+1$ dimensions, we introduce measurements with a spatiotemporally-modulated density. Exploiting the known critical properties of the measurement-induced entanglement transition, this allows us to engineer arbitrary geometries for the bulk space (with a fixed topology). These geometries in turn control the entanglement structure of the boundary (output) state. We demonstrate our approach by giving concrete protocols for two geometries of interest in two dimensions: the hyperbolic half-plane and a spatial section of the BTZ black hole. We numerically verify signatures of the underlying entanglement geometry, including a direct imaging of entanglement wedges by using locally-entangled reference qubits. Our results provide a concrete platform for realizing geometric entanglement structures on near-term quantum simulators.
\end{abstract}

\maketitle


{\it Introduction.} 
Entanglement is a fundamental unifying concept across the domains of many-body physics, quantum information science and gravity. It plays an important role in our understanding of equilibrium phases of matter\cite{calabrese_entanglement_2004,levin2006detecting,kitaev2006topological}, nonequilibrium phenomena such as thermalization~\cite{calabrese_evolution_2005,abanin_colloquium_2019}, and the holographic principle~\cite{ryu_aspects_2006}. 
Quantum circuits---models of dynamics composed of discrete, few-body unitary interactions---have emerged as powerful toy models for exploring these fundamental ideas, while also enabling concrete experimental connections to recently-developed digital quantum simulators and computers~\cite{hosur_chaos_2016,nahum_quantum_2017,bertini_exact_2018,nahum_operator_2018,khemani_operator_2018,arute_quantum_2019,zhou_emergent_2019,piroli_random_2020,mi_information_2021,potter_entanglement_2022,fisher_random_2022}.

In recent years, novel types of non-equilibrium phases of matter defined by entanglement have been identified in models of {\it monitored dynamics}, where unitary interactions are interspersed with measurements~\cite{skinner_measurement-induced_2019,li_quantum_2018,li_measurement-driven_2019,gullans_dynamical_2020,choi_quantum_2020}. 
The most well-known manifestation of this phenomenon is the measurement-induced entanglement transition (MIPT) where, as a function of the spatiotemporal density of measurements $\rho \in[0,1]$, late-time states of the dynamics may show sharply different structures of entanglement---``area-law''~\cite{eisert_arealaw_2010} or ``volume-law''~\cite{page_entropy_1993}---separated by a critical point $\rho = \rho_c$ described by a conformal field theory (CFT)~\cite{jian_measurement-induced_2020,bao_theory_2020,li_conformal_2021}. 

A notable aspect of the volume-law phase in these models is its geometric description. Namely the entanglement of a subsystem $A$ in the output (boundary) state is related to the area of a {\it minimal membrane} that bounds region $A$. Here we will focus on system in dimension 1+1, where such membrane is a curve $\gamma$ also known as the entanglement domain wall~\cite{nahum_quantum_2017,jonay_coarse-grained_2018,li_statistical_2021,li_entanglement_2023}.
This description is analogous to the Ryu-Takayanagi surface in holographic duality~\cite{ryu_aspects_2006}, and in particular to its realization in {\it holographic tensor networks} which are important toy models for the holographic duality~\cite{pastawski_holographic_2015,hayden2016holographic,jahn2021holographic}. 
Given this connection, it is interesting to ask whether suitably-designed models of monitored dynamics may reproduce interesting features of holographic or gravitational theories, and possibly bring them within reach of near-term quantum simulators.

In this Letter, we show that this is the case by providing a general recipe for engineering a target entanglement geometry by means of monitored quantum circuits with a spatiotemporally-modulated density of measurements. In other words, given a target Euclidian metric $g_{ij}(x,t)$ 
we give a density of measurements $\rho(x,t)$ which, inserted in a suitable family of circuits, gives the desired bulk geometry.
In particular, in output states of the circuit the entanglement entropy $S_A$ of any interval $A$ is proportional to the geodesic distance---according to the target metric $g(x,t)$---between the interval's endpoints. Other gravitational features, e.g. black hole horizons, can be represented by initial states or suitable boundary conditions. 

We illustrate our proposal by focusing on two interesting examples of 2D metrics that arise in gravity: the Poincare patch of a two-dimensional hyperbolic space, 
and the metric of a spatial section of a Bañados, Teitelboim and Zanelli (BTZ) black hole~\cite{banados_black_1992}. By numerical simulations of Clifford circuit models, we show several distinctive signatures of the underlying entanglement geometry. 
These include a direct imaging of the entanglement wedge of a subsystem $A$, obtained from the mutual information between $A$ and a reference qubit $R$ entangled at a variable space-time location $(x,t)$ during the dynamics.

Our results introduce new classes of monitored entanglement structures, including logarithmically-entangled states generated by the Euclidean anti de-Sitter (AdS) metric. Unlike the critical states generated by the dynamics at $\rho = \rho_c$, the logarithmic scaling of entanglement is not due to a CFT but rather to a minimal-membrane picture, much like that in the volume-law phase; however the length of AdS geodesics yields $S_A \sim \ln|A|$ instead of the flat metric result $S_A \sim |A|$.
Moreover, since our method is based on Clifford circuits, these structures may be accessible on digital quantum simulators supplemented with (polynomial-time) classical simulation~\cite{aaronson_improved_2004}, paving the way for the realization of geometric entanglement structures in realistic experiments.

\begin{figure}
    \centering
    \includegraphics[width=.8\columnwidth]{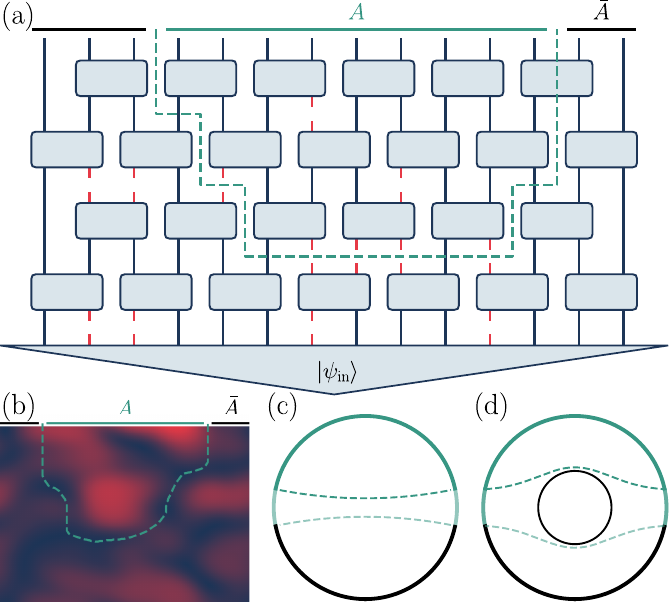}
    \caption{Visualizing the experimental setup and geodesics. Subfigure (a) shows a dual unitary circuit with random measurements (red) as a function of spacetime and an minimal cut (turquoise). Notice that boundary conditions are important for the BTZ case, but not for the AdS case in which, in principle, has an infinitely deep circuit. In subfigure (b) we show how a large circuit approximates a continuous spacetime with a geodesic. Subfigures (c) and (d) show entanglement geodesics for AdS and a constant-time slice of BTZ respectively, corresponding to states on the boundary in the Poincar\'e disk model. Notice that the entanglement geodesics vary continuously with size of the subregion in the AdS case while they jump in the BTZ case due to the presence of a horizon.} 
    \label{fig:experimental_picture}
\end{figure}

{\it Geometry from measurement.}
We consider 1+1D circuits on $L$-qubit chains with periodic boundary conditions. The circuits are made of random 2-qubit unitary gates in a brickwork pattern and random single-qubit measurement of qubit $x$ at time $t$ with probability $\rho(x,t)$, slowly varying with position $x$ and time $t$. The circuit depth and the initial state are to be specified in each case; we aim to characterize the entanglement structure of the final state. The setup is sketched in \cref{fig:experimental_picture}. 

With a uniform measurement rate $\rho$, prior works\cite{skinner_measurement-induced_2019,li_quantum_2018,li_measurement-driven_2019} have shown that there is a phase transition in the final state's entanglement, from an area-law to a volume-law, at a critical measurement rate $0<\rho_c<1$. 
The entanglement computation maps onto the statistical mechanics of an effective spin model, where the ``spin'' variables describe the pairing of wavefunction replicas~\cite{jian_measurement-induced_2020,bao_theory_2020,choi_quantum_2020,potter_entanglement_2022}; the entanglement transition corresponds to an ordering transition in this magnet. Specifically, the entropy $S_A$ of a region $A$ in the circuit maps onto the free energy cost of flipping boundary conditions in $A$ in the magnet: in the ordered phase, this boundary twist seeds a domain that extends into the bulk~\cite{vasseur_entanglement_2019,bao_theory_2020,li_statistical_2021,li_entanglement_2023}. The domain wall carries a finite line tension, giving volume-law entanglement entropy $S_A \sim s|A|$, with $s$ the the entropy density.

The entropy density $s(\rho)$ in the vicinity of the critical point scales as $s(\rho) \sim (\rho_c - \rho)^\nu$, reflecting a divergent correlation length $\xi \sim |\rho_c-\rho|^{-\nu}$ in the stat-mech model~\cite{gullans_dynamical_2020}. In the model we consider, the correlation length exponent is known to be $\nu \simeq 1.28$~\cite{zabalo_critical_2020}.
Our goal is to leverage this predictable scaling form to design a spacetime-dependent measurement rate $\rho(x, t)$ that induces an effective metric $g(x, t)$ in the bulk of the circuit, in the sense that the entanglement domain wall for any given subsystem $A$ is given by a geodesic of $g$ connecting the endpoints of $A$.

If the measurement rate varies sufficiently slowly with $x$ and $t$, then upon coarse-graining we have regions of spacetime where the magnet is locally at equilibrium with respect to the measurement rate $\rho(x,t)$. 
Then domain walls through that region have a well-defined line tension. The free energy cost $\diff F$ of a differential element $\diff r = (\diff x, \diff t)$ of the domain wall is 
\begin{equation}
    \diff F^2 
    \sim g_{ij} \diff r_i \diff r_j 
    = g_{xx} \diff x^2 + g_{tt} \diff t^2
    \label{eq:metric_F}
\end{equation}
where the second step assumes spatial and/or temporal inversion symmetry.
By identifying the free energy element $\diff F$ in the magnet with the entropy element $\diff S$ in the circuit, we obtain
\begin{equation}
    \diff S^2 
    = s^2 (\diff x^2 + v_E^2 \diff t^2)
    \label{eq:metric}
\end{equation}
where the functions $s = \sqrt{g_{xx}}$ and $v_E = \sqrt{g_{tt} / g_{xx}}$ play the roles of local ``entropy density'' and ``entanglement velocity'', respectively\footnote{In the sense that, for a uniform metric, one has $S_A \sim s|A|$ for an interval in a late-time state (geodesic directed along $\hat{x}$) and $S \sim sv_E \Delta t$ for a half-cut in a product state evolved for a short time $\Delta t$ (geodesic directed along $\hat{t}$).}.

Let us first consider the case of an isotropic metric $g_{ij}$, 
\begin{equation}
    \diff S^2 = s^2(x,t) (\diff x^2 + \diff t^2).
    \label{eq:isotropic}
\end{equation}
Metrics of this form can be systematically realized by restricting the gates in the brickwork circuit to be {\it dual-unitary}, i.e. unitary in both the space and time directions~\cite{bertini_exact_2018,bertini_exact_2019}~\footnote{The fact that $v_E=1$ in these circuits was previously exploited to simplify the extraction of critical exponents at the transition~\cite{zabalo_operator_2022}.}. 
As already mentioned, near the critical point ($\rho \lesssim \rho_c$) the entropy density has a predicted scaling $s \sim (\rho_c-\rho)^{\nu}$, therefore an assignment of $\rho(x,t)$ that produces a target isotropic metric $s^2(x,t)(\diff x^2 + \diff t^2)$ is given by 
\begin{equation}
    \rho(x, t) = \rho_c -  \kappa s(x,t)^{1/\nu}
    \label{eq:measurement_rate}
\end{equation}
where $\kappa$ is a positive constant. 

More general metrics can be reduced to the isotropic form by a suitable diffeomorphism $(x,t)\mapsto (\tilde{x}(x,t),\tilde{t}(x,t))$. Crucially however, $\tilde{x}$ must act trivially on the $t = 0$ slice, i.e., the distance between points in the final state is a physical, invariant property that we are not allowed to redefine (otherwise the notion of entanglement scaling with subsystem size loses meaning). 
For metrics with space and/or time-reversal symmetry, as in \cref{eq:metric}, this can be achieved by a rescaling of the time coordinate only such that $\partial_t \tilde{t} = v_E(x,t)$. This manifestly preserves spatial distance in the final state.

In summary, given a target metric $g(x,t)$ [\cref{eq:metric}], our prescription is as follows: 
(i) reparametrize time to make the metric isotropic [\cref{eq:isotropic}]; 
(ii) obtain the measurement rate $\rho(x,t)$ [\cref{eq:measurement_rate}]; 
(iii) run brickwork circuits with dual-unitary gates and single-qubit measurement density $\rho(x,t)$ for $t \in [-T,0]$ (the circuit depth $T$ and initial state vary by case and are discussed below). 
This generates ensembles of output states on the $t = 0$ surface whose entanglement has a geometric description determined by the target metric.

{\it Two testbeds.}
We test this procedure on two metrics that are of interest in gravity. 
The first is the Poincare AdS metric
\begin{equation}
    \diff \ell^2 = \frac{\diff x^2 + \diff t^2}{t^2}.
    \label{eq:AdS}
\end{equation}
with $t < 0 $. $x$ is compactified into a circle with period $L$.
This is already in the isotropic form, with $s(x,t) = 1/|t|$ and thus $\rho(x,t) = \rho_c - \kappa |t|^{-1/\nu}$. 
This corresponds to starting the dynamics (at $t = - \infty$) at the critical point $\rho = \rho_c$, and then gradually lowering $\rho$ into the volume-law phase. 
$l \equiv \kappa^\nu$ is the AdS radius.
Note that as the metric diverges for $t \to 0^-$, the required measurement density becomes negative; we stop the evolution when $\rho = 0$. This corresponds to placing the state at a finite radius rather than on the asymptotic boundary of the space. 

The second metric represents a spatial slice of the (2+1)-dimensional BTZ metric~\cite{banados_black_1992}, given by
\begin{equation}
    \diff \ell^2 = \frac{l^2 \diff r^2}{r^2 - r_h^2} + r^2 \diff \phi^2
    \label{eq:BTZ}
\end{equation}
where $r$ and $\phi$ are radial and angular coordinates respectively, $r_h$ is the radius of the black hole horizon (we will focus on the non-rotating BTZ black holes with a single horizon)
and $l$ is the AdS radius.
We map the angular coordinate to position, $x \equiv L\phi / 2\pi$ ($L$ is the number of qubits in the chain with periodic boundary conditions), and the radial coordinate to time via 
\begin{equation}
    t \equiv T \left[\frac{2}{\pi} \arctan \left(\sqrt{(r / r_h)^2 -1} \right) -1 \right], 
    \label{eq:t_reparametrize}
\end{equation}
with $T \equiv Ll/4r_h$ being the circuit depth ($-T < t < 0$). The initial time $t = -T$ corresponds to the horizon ($r = r_h$), while the final time $t \to 0$ corresponds to the boundary ($r \to \infty$). The aspect ratio of the space is fixed by the size of the black hole horizon in units of the AdS radius: $L/T = 4r_h/l$. 

The mapping in \cref{eq:t_reparametrize} reduces the metric to the isotropic form [\cref{eq:isotropic}] with entropy density
\begin{equation}
    s(x, t) = \frac{\pi l}{2T} \csc \left(\frac{\pi |t|}{2T}\right).
\end{equation}
For $t\to 0^-$ this reduces to the AdS case, $s \simeq l/|t|$. 
However unlike the AdS case, where the initial state is infinitely far in the past and thus irrelevant (due to purification~\cite{gullans_dynamical_2020}), in the BTZ case the initial state is a time $T = O(L)$ in the past and plays an important role as we will see below. 


{\it Numerical verification.}
We begin with a direct comparison of the entanglement entropy $S_A$ of contiguous subsystems in the final state to the lengths of the geodesic $\gamma(A)$ connecting the subsystem's endpoints. 
We numerically simulate 1D brickwork circuits of dual-unitary Clifford gates, interspersed with single-qubit Pauli measurements with probability $\rho(x,t)$ [\cref{eq:measurement_rate}] for the AdS and BTZ metrics. 
These circuits can be simulated efficiently by the stabilizer method~\cite{aaronson_improved_2004} and the entanglement transition occurs at $\rho_c = 0.2050(5)$, see Ref.~\cite{zabalo_operator_2022} and Supplementary Material (SM)~\footnote{See online Supplementary Material for numerical study of the entanglement transition, derivation of the BTZ measurement rate and geodesic length, and numerics on AdS states.}.

\begin{figure}
    \includegraphics[width=.9\linewidth]{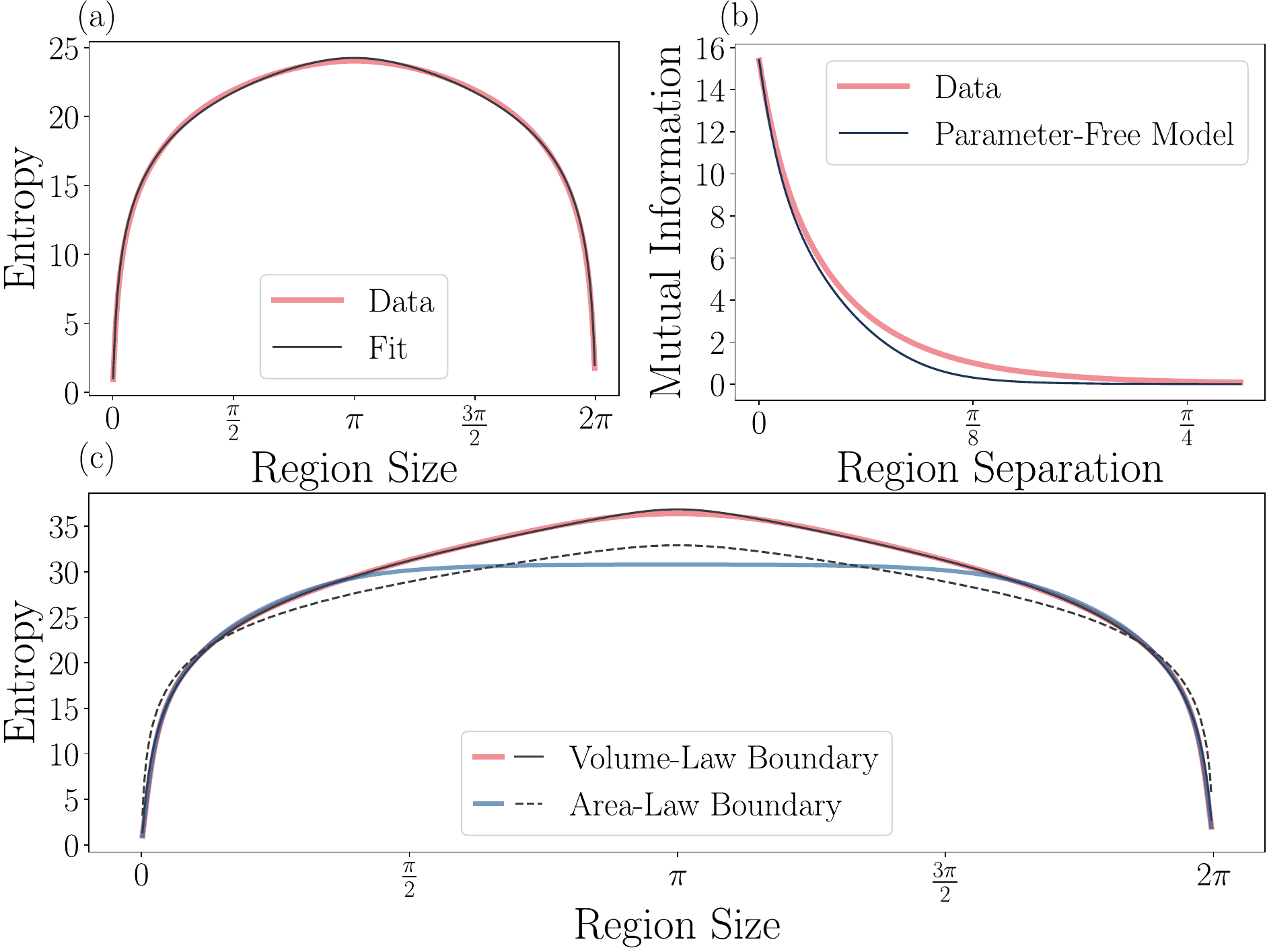}
    \caption{Analyzing the relationship between entanglement entropies and geodesic lengths. Subfigures (a) and (c) show the entanglement entropy of a single interval of varying size on the boundary, such as those shown in \ref{fig:experimental_picture}. Subfigure (a) shows the entanglement entropy of a region on the boundary of hyperbolic space, while subfigure (c) shows the entanglement entropy of a region on the boundary of a space containing a BTZ black hole with a product state (blue, dotted) or a maximal volume-law state (pink, solid) placed on the horizon. In subfigure (b) we depict the mutual information between two regions of size $\frac{\pi}{4}$ as a function of their separation. No free parameters are needed to establish the mutual information model, as the entropy per unit length is derived from the fit in subfigure (a).}
    \label{fig:simple_comparison}
\end{figure}

Numerical results for the entanglement entropy $S_A$ are shown in \cref{fig:simple_comparison} against analytical expressions for the length of the geodesic $\gamma(A)$. 
In comparing the two, one must take into account the UV-regularization induced by the lattice spacing and circuit time step, which is done by including appropriate fit parameters in the expression for the geodesic distance.
Each panel of \cref{fig:simple_comparison} shows the average over circuit realizations of the entanglement entropy $S_A$ of the final state as a function of the size (number of qubits) of region $A$. For the AdS metric, \cref{fig:simple_comparison}(a), we find good agreement with the prediction based on the geodesic length in hyperbolic space, $S_A \sim \ln|A|$ (note that while in principle the dynamics starts at $t = -\infty$, we find well-converged results starting at $t = -L$). 
This quantitative agreement demonstrates the robustness of this state-preparation procedure.

We note that this family of logarithmically-entangled states is different from the critical states that arise at $\rho = \rho_c$. In particular the ``central charge'' $c$ in $S_A \sim \frac{c}{3} \ln|A|$ (which is universal for critical states at $\rho = \rho_c$) is predicted to be proportional to the AdS radius, $c \propto l$. We verify numerically (see SM~\cite{Note3}) that $c \simeq a l + b$, where the constant offset $b$ comes from the background critical state ($l \to 0$ yields $\rho \to \rho_c$). 

Moving on to the BTZ metric, \cref{fig:simple_comparison}(b) considers a product initial state at $t = -T$, and shows a substantial disagreement with the prediction. In particular the entanglement entropy plateaus at a subsystem size $|A| < L/2$ rather than displaying the expected cusp at $L/2$. This happens because, with a disentangled initial state, it becomes convenient for the geodesic to exit through the $t = -T$ boundary of the circuit. This causes $S_A$ to saturate when the entanglement domain wall splits into two disjoint pieces on either side of $A$. 
To resolve this problem we instead use a highly entangled initial state (prepared by a random unitary circuit of depth $2L$), \cref{fig:simple_comparison}(d), which gives excellent agreement with the analytical prediction. 
This corresponds to the fact that the entropy density of a black hole is the maximum entropy density allowed on a surface due to the Bekenstein bound \cite{bekenstein1981universal}. Geometrically this forces the entanglement geodesic to remain within the circuit, giving the length we expect based on the BTZ geometry.

{\it Transitions in the entanglement wedge.}
As noted above, a distinctive feature of the geometric picture is that it predicts sharp transitions in the entropy of subsystems~\cite{headrick2010entanglement} 
(seen as cusps in $S_A$ vs $|A|$ in \cref{fig:simple_comparison}). 
Consider two disjoint intervals $A$ and $B$ on the boundary. Their mutual information $I(A:B)$ involves the joint entropy $S_{AB}$, whose minimal surface can take two inequivalent configurations: $\gamma(A) \cup \gamma(B)$, or $\gamma(C) \cup \gamma(ACB)$, where $C$ is the shortest interval between $A$ and $B$. The former gives $S_{AB}=S_A + S_B$ and thus $I(A:B) = 0$, the latter can give nontrivial mutual information. 
Thus we expect a transition when the lengths of the two configurations are equal.~\cite{li_statistical_2021}.

Having already determined the entropy per unit length of geodesic from fitting to the single-region case, we can now do a parameter-free comparison between the numerically-computed $I(A:B)$ and the analytically derived minimal geodesic length.
If the geodesic lengths of the two configurations $S_1$ and $S_2$ are very different, we have $S \simeq \min(S_1, S_2)$. If they are comparable, we must take into account both saddles, yielding $S \simeq -\log \left( e^{-S_1} + e^{-S_2} \right)$. 
Results for the mutual information between two regions of equal size as a function of their separation are shown in \cref{fig:simple_comparison}(d). 
There is again close agreement between the analytical prediction and the numerics, though the geometric analysis predicts a more sharply decreasing mutual information than that seen in the numerics. This is likely due to fluctuations around the saddle points near the transition which are not taken into account in the RT formula. Nonetheless the overall agreement shows that the geometric perspective holds also beyond single intervals.

We can go beyond the indirect diagnostics used above, and directly image the entanglement wedge of arbitrary subsystems by entangling additional ``reference qubits'' at specific space-time points $(x,t)$ during the state-preparation process~\cite{gullans_scalable_2020,ippoliti_entanglement_2021}~\footnote{As a practical matter in simulations, after performing a measurement on a site, we may entangle the qubit at that location with a reference $R$ instead of leaving it in the measured state.}. Then, the mutual information between the reference and a subsystem $A$ of the final state can be used to determine the entanglement wedge. Indeed, if the point $(x,t)$ where the reference is entangled lies inside the entanglement wedge of $A$, then the operators $X_R \equiv X_{(x,t)}$ and $Z_R \equiv Z_{(x,t)}$ can be reconstructed from $A$ with finite probability~\footnote{It is always possible that the location where $R$ is entangled gets measured again immediately afterwards, disentangling $R$ with some probability even inside the wedge.}, implying $I(A:R) > 0$. If $(x,t)$ is outside the wedge, then the operators cannot be reconstructed, and $I(A:R) = 0$. 

\begin{figure}
    \centering
    \includegraphics[width=\columnwidth]{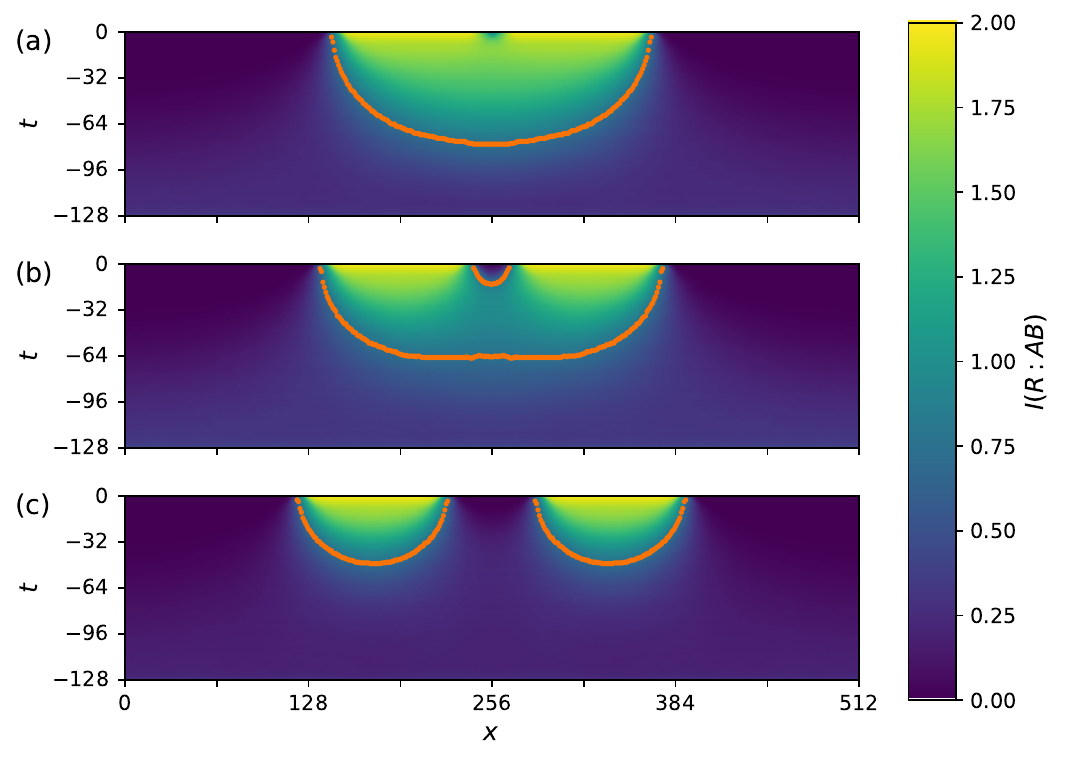}
    \caption{Imaging the entanglement wedge of two intervals in the BTZ metric. 
    Color plots show mutual information $I(R:AB)$ between a reference qubit $R$ entangled at space-time location $(x,t)$ and the union of two intervals $A$, $B$. The intervals contain $102$ qubits each and are separated by (a) $\delta x = 16$, (b) $\delta x = 32$, (c) $\delta x = 64$ qubits. 
    Orange lines denote contours with $I = 0.75$, clearly showing a transition in the structure of the entanglement wedge as a function of separation $\delta x$.
    The system size is $L = 512$ and we set $r_h = l = 0.5$, giving $T=128$. Data averaged over $4\times 10^3$ realizations of Clifford circuits.
    }
    \label{fig:entanglement_wedges}
\end{figure}

Entanglement wedges imaged with this technique for the BTZ metric are shown in \cref{fig:entanglement_wedges}. Despite the presence of fluctuations which blur the contours of the entanglement wedge, the crossover between the two possible geodesic configurations is clearly visible as the two regions $A$, $B$ are moved further apart. 
When they are close together they share the same wedge, when they are far away the wedge separates into disconnected components, and near the crossover point both configurations are visible, demonstrating the effects of both saddle points.

{\it Discussion.} 
We have introduced a method for constructing ensembles of states with holographic entanglement structure whose bulk geometry can be arbitrarily engineered.
The method uses 1D arrays of qubits with local couplings, making it well-suited to near term quantum computer architectures. The protocol is dynamical, with time playing the role of a bulk dimension, and boundary states produced at the final time. The bulk geometry is induced by a spatiotemporally-modulated density of projective measurements $\rho(x,t)$.
By examining the AdS metric as a testbed of our proposal, we have identified a new class of logarithmically-entangled output states of monitored dynamics, illustrating the new possibilities offered by spatiotemporal modulations of the measurement rate~\cite{zabalo_infinite-randomness_2022}.

A practical limitation of our proposal is that the desired entanglement structure is lost when averaging over measurement outcomes. Post-selection of outcomes incurs an exponential sampling overhead.
One way around this issue is to use adaptive circuits which contain feedback operations to correct any ``unwanted'' measurement outcomes (generally via a global unitary) and drive the system to a desired trajectory. In Clifford circuits with Pauli measurements, such corrective operations can be found with efficient classical algorithms~\cite{gullans_scalable_2020}.
Another approach is to view our proposal as a classical algorithm for generating ensembles of stabilizer states with the desired holographic properties; such stabilizer states can then be compiled, by a polynomial-time algorithm, into unitary Clifford circuits of depth $O(N)$ acting on a fiduciary state such as $|0\rangle^{\otimes N}$. The unitary circuits can then be straightforwardly implemented on quantum hardware. 

This method is hardware-efficient relative to other toy models of holography based on discrete qubit arrays~\cite{pastawski_holographic_2015,hayden2016holographic,jahn2021holographic} since it does not require the geometry to be hard-coded into the hardware connectivity---nearest-neighbor interactions in flat space are sufficient. 
It is also qubit-efficient since it takes advantage of time as an extra dimension, so no extra qubits are needed to model the bulk (all physical qubits contribute to the boundary state). 
However on noisy hardware, finite coherence would practically limit the evolution time and thus the size of realizable geometries.


In this work we have focused on a saddle-point analysis, taking entanglement to be represented exactly by the size of the minimal surface. 
While this is borne out with good accuracy in numerics, it is known that fluctuations around the saddle point play an important role~\cite{li_entanglement_2023,weinstein_measurement-induced_2022}. An interesting goal for future work is to systematically analyze such fluctuations beyond the flat metric case. 

\noindent{\bf Acknowledgement.} 
MI acknowledges discussions with Vedika Khemani.
This work is supported by the National Science Foundation under grant No.2111998 (AC and XLQ), the Simons Foundation (XLQ), and the Gordon and Betty Moore Foundation’s EPiQS Initiative through Grant GBMF8686 (MI). This work is also supported in part by the DOE Office of Science, Office of High Energy Physics, the grant de-sc0019380. 
Numerical calculations were carried out in part on Stanford Research Computing's Sherlock cluster. 

\bibliography{spacetimemeasured}

\begin{thebibliography}{51}%
\makeatletter
\providecommand \@ifxundefined [1]{%
 \@ifx{#1\undefined}
}%
\providecommand \@ifnum [1]{%
 \ifnum #1\expandafter \@firstoftwo
 \else \expandafter \@secondoftwo
 \fi
}%
\providecommand \@ifx [1]{%
 \ifx #1\expandafter \@firstoftwo
 \else \expandafter \@secondoftwo
 \fi
}%
\providecommand \natexlab [1]{#1}%
\providecommand \enquote  [1]{``#1''}%
\providecommand \bibnamefont  [1]{#1}%
\providecommand \bibfnamefont [1]{#1}%
\providecommand \citenamefont [1]{#1}%
\providecommand \href@noop [0]{\@secondoftwo}%
\providecommand \href [0]{\begingroup \@sanitize@url \@href}%
\providecommand \@href[1]{\@@startlink{#1}\@@href}%
\providecommand \@@href[1]{\endgroup#1\@@endlink}%
\providecommand \@sanitize@url [0]{\catcode `\\12\catcode `\$12\catcode
  `\&12\catcode `\#12\catcode `\^12\catcode `\_12\catcode `\%12\relax}%
\providecommand \@@startlink[1]{}%
\providecommand \@@endlink[0]{}%
\providecommand \url  [0]{\begingroup\@sanitize@url \@url }%
\providecommand \@url [1]{\endgroup\@href {#1}{\urlprefix }}%
\providecommand \urlprefix  [0]{URL }%
\providecommand \Eprint [0]{\href }%
\providecommand \doibase [0]{https://doi.org/}%
\providecommand \selectlanguage [0]{\@gobble}%
\providecommand \bibinfo  [0]{\@secondoftwo}%
\providecommand \bibfield  [0]{\@secondoftwo}%
\providecommand \translation [1]{[#1]}%
\providecommand \BibitemOpen [0]{}%
\providecommand \bibitemStop [0]{}%
\providecommand \bibitemNoStop [0]{.\EOS\space}%
\providecommand \EOS [0]{\spacefactor3000\relax}%
\providecommand \BibitemShut  [1]{\csname bibitem#1\endcsname}%
\let\auto@bib@innerbib\@empty
\bibitem [{\citenamefont {Calabrese}\ and\ \citenamefont
  {Cardy}(2004)}]{calabrese_entanglement_2004}%
  \BibitemOpen
  \bibfield  {author} {\bibinfo {author} {\bibfnamefont {P.}~\bibnamefont
  {Calabrese}}\ and\ \bibinfo {author} {\bibfnamefont {J.}~\bibnamefont
  {Cardy}},\ }\bibfield  {title} {\bibinfo {title} {Entanglement entropy and
  quantum field theory},\ }\href
  {https://doi.org/10.1088/1742-5468/2004/06/P06002} {\bibfield  {journal}
  {\bibinfo  {journal} {Journal of Statistical Mechanics: Theory and
  Experiment}\ }\textbf {\bibinfo {volume} {2004}},\ \bibinfo {pages} {P06002}
  (\bibinfo {year} {2004})}\BibitemShut {NoStop}%
\bibitem [{\citenamefont {Levin}\ and\ \citenamefont
  {Wen}(2006)}]{levin2006detecting}%
  \BibitemOpen
  \bibfield  {author} {\bibinfo {author} {\bibfnamefont {M.}~\bibnamefont
  {Levin}}\ and\ \bibinfo {author} {\bibfnamefont {X.-G.}\ \bibnamefont
  {Wen}},\ }\bibfield  {title} {\bibinfo {title} {Detecting topological order
  in a ground state wave function},\ }\href@noop {} {\bibfield  {journal}
  {\bibinfo  {journal} {Physical review letters}\ }\textbf {\bibinfo {volume}
  {96}},\ \bibinfo {pages} {110405} (\bibinfo {year} {2006})}\BibitemShut
  {NoStop}%
\bibitem [{\citenamefont {Kitaev}\ and\ \citenamefont
  {Preskill}(2006)}]{kitaev2006topological}%
  \BibitemOpen
  \bibfield  {author} {\bibinfo {author} {\bibfnamefont {A.}~\bibnamefont
  {Kitaev}}\ and\ \bibinfo {author} {\bibfnamefont {J.}~\bibnamefont
  {Preskill}},\ }\bibfield  {title} {\bibinfo {title} {Topological entanglement
  entropy},\ }\href@noop {} {\bibfield  {journal} {\bibinfo  {journal}
  {Physical review letters}\ }\textbf {\bibinfo {volume} {96}},\ \bibinfo
  {pages} {110404} (\bibinfo {year} {2006})}\BibitemShut {NoStop}%
\bibitem [{\citenamefont {Calabrese}\ and\ \citenamefont
  {Cardy}(2005)}]{calabrese_evolution_2005}%
  \BibitemOpen
  \bibfield  {author} {\bibinfo {author} {\bibfnamefont {P.}~\bibnamefont
  {Calabrese}}\ and\ \bibinfo {author} {\bibfnamefont {J.}~\bibnamefont
  {Cardy}},\ }\bibfield  {title} {\bibinfo {title} {Evolution of entanglement
  entropy in one-dimensional systems},\ }\href
  {https://doi.org/10.1088/1742-5468/2005/04/P04010} {\bibfield  {journal}
  {\bibinfo  {journal} {Journal of Statistical Mechanics: Theory and
  Experiment}\ }\textbf {\bibinfo {volume} {2005}},\ \bibinfo {pages} {P04010}
  (\bibinfo {year} {2005})}\BibitemShut {NoStop}%
\bibitem [{\citenamefont {Abanin}\ \emph {et~al.}(2019)\citenamefont {Abanin},
  \citenamefont {Altman}, \citenamefont {Bloch},\ and\ \citenamefont
  {Serbyn}}]{abanin_colloquium_2019}%
  \BibitemOpen
  \bibfield  {author} {\bibinfo {author} {\bibfnamefont {D.~A.}\ \bibnamefont
  {Abanin}}, \bibinfo {author} {\bibfnamefont {E.}~\bibnamefont {Altman}},
  \bibinfo {author} {\bibfnamefont {I.}~\bibnamefont {Bloch}},\ and\ \bibinfo
  {author} {\bibfnamefont {M.}~\bibnamefont {Serbyn}},\ }\bibfield  {title}
  {\bibinfo {title} {Colloquium: Many-body localization, thermalization, and
  entanglement},\ }\href {https://doi.org/10.1103/RevModPhys.91.021001}
  {\bibfield  {journal} {\bibinfo  {journal} {Rev. Mod. Phys.}\ }\textbf
  {\bibinfo {volume} {91}},\ \bibinfo {pages} {021001} (\bibinfo {year}
  {2019})}\BibitemShut {NoStop}%
\bibitem [{\citenamefont {Ryu}\ and\ \citenamefont
  {Takayanagi}(2006)}]{ryu_aspects_2006}%
  \BibitemOpen
  \bibfield  {author} {\bibinfo {author} {\bibfnamefont {S.}~\bibnamefont
  {Ryu}}\ and\ \bibinfo {author} {\bibfnamefont {T.}~\bibnamefont
  {Takayanagi}},\ }\bibfield  {title} {\bibinfo {title} {Aspects of holographic
  entanglement entropy},\ }\href
  {https://doi.org/10.1088/1126-6708/2006/08/045} {\bibfield  {journal}
  {\bibinfo  {journal} {Journal of High Energy Physics}\ }\textbf {\bibinfo
  {volume} {2006}},\ \bibinfo {pages} {045} (\bibinfo {year}
  {2006})}\BibitemShut {NoStop}%
\bibitem [{\citenamefont {Hosur}\ \emph {et~al.}(2016)\citenamefont {Hosur},
  \citenamefont {Qi}, \citenamefont {Roberts},\ and\ \citenamefont
  {Yoshida}}]{hosur_chaos_2016}%
  \BibitemOpen
  \bibfield  {author} {\bibinfo {author} {\bibfnamefont {P.}~\bibnamefont
  {Hosur}}, \bibinfo {author} {\bibfnamefont {X.-L.}\ \bibnamefont {Qi}},
  \bibinfo {author} {\bibfnamefont {D.~A.}\ \bibnamefont {Roberts}},\ and\
  \bibinfo {author} {\bibfnamefont {B.}~\bibnamefont {Yoshida}},\ }\bibfield
  {title} {\bibinfo {title} {Chaos in quantum channels},\ }\href
  {https://doi.org/10.1007/JHEP02(2016)004} {\bibfield  {journal} {\bibinfo
  {journal} {Journal of High Energy Physics}\ }\textbf {\bibinfo {volume}
  {2016}},\ \bibinfo {pages} {4} (\bibinfo {year} {2016})}\BibitemShut
  {NoStop}%
\bibitem [{\citenamefont {Nahum}\ \emph {et~al.}(2017)\citenamefont {Nahum},
  \citenamefont {Ruhman}, \citenamefont {Vijay},\ and\ \citenamefont
  {Haah}}]{nahum_quantum_2017}%
  \BibitemOpen
  \bibfield  {author} {\bibinfo {author} {\bibfnamefont {A.}~\bibnamefont
  {Nahum}}, \bibinfo {author} {\bibfnamefont {J.}~\bibnamefont {Ruhman}},
  \bibinfo {author} {\bibfnamefont {S.}~\bibnamefont {Vijay}},\ and\ \bibinfo
  {author} {\bibfnamefont {J.}~\bibnamefont {Haah}},\ }\bibfield  {title}
  {\bibinfo {title} {Quantum {Entanglement} {Growth} under {Random} {Unitary}
  {Dynamics}},\ }\href {https://doi.org/10.1103/PhysRevX.7.031016} {\bibfield
  {journal} {\bibinfo  {journal} {Physical Review X}\ }\textbf {\bibinfo
  {volume} {7}},\ \bibinfo {pages} {031016} (\bibinfo {year}
  {2017})}\BibitemShut {NoStop}%
\bibitem [{\citenamefont {Bertini}\ \emph {et~al.}(2018)\citenamefont
  {Bertini}, \citenamefont {Kos},\ and\ \citenamefont
  {Prosen}}]{bertini_exact_2018}%
  \BibitemOpen
  \bibfield  {author} {\bibinfo {author} {\bibfnamefont {B.}~\bibnamefont
  {Bertini}}, \bibinfo {author} {\bibfnamefont {P.}~\bibnamefont {Kos}},\ and\
  \bibinfo {author} {\bibfnamefont {T.}~\bibnamefont {Prosen}},\ }\bibfield
  {title} {\bibinfo {title} {Exact {Spectral} {Form} {Factor} in a {Minimal}
  {Model} of {Many}-{Body} {Quantum} {Chaos}},\ }\href
  {https://doi.org/10.1103/PhysRevLett.121.264101} {\bibfield  {journal}
  {\bibinfo  {journal} {Physical Review Letters}\ }\textbf {\bibinfo {volume}
  {121}},\ \bibinfo {pages} {264101} (\bibinfo {year} {2018})}\BibitemShut
  {NoStop}%
\bibitem [{\citenamefont {Nahum}\ \emph {et~al.}(2018)\citenamefont {Nahum},
  \citenamefont {Vijay},\ and\ \citenamefont {Haah}}]{nahum_operator_2018}%
  \BibitemOpen
  \bibfield  {author} {\bibinfo {author} {\bibfnamefont {A.}~\bibnamefont
  {Nahum}}, \bibinfo {author} {\bibfnamefont {S.}~\bibnamefont {Vijay}},\ and\
  \bibinfo {author} {\bibfnamefont {J.}~\bibnamefont {Haah}},\ }\bibfield
  {title} {\bibinfo {title} {Operator {Spreading} in {Random} {Unitary}
  {Circuits}},\ }\href {https://doi.org/10.1103/PhysRevX.8.021014} {\bibfield
  {journal} {\bibinfo  {journal} {Physical Review X}\ }\textbf {\bibinfo
  {volume} {8}},\ \bibinfo {pages} {021014} (\bibinfo {year}
  {2018})}\BibitemShut {NoStop}%
\bibitem [{\citenamefont {Khemani}\ \emph {et~al.}(2018)\citenamefont
  {Khemani}, \citenamefont {Vishwanath},\ and\ \citenamefont
  {Huse}}]{khemani_operator_2018}%
  \BibitemOpen
  \bibfield  {author} {\bibinfo {author} {\bibfnamefont {V.}~\bibnamefont
  {Khemani}}, \bibinfo {author} {\bibfnamefont {A.}~\bibnamefont
  {Vishwanath}},\ and\ \bibinfo {author} {\bibfnamefont {D.~A.}\ \bibnamefont
  {Huse}},\ }\bibfield  {title} {\bibinfo {title} {Operator {Spreading} and the
  {Emergence} of {Dissipative} {Hydrodynamics} under {Unitary} {Evolution} with
  {Conservation} {Laws}},\ }\href {https://doi.org/10.1103/PhysRevX.8.031057}
  {\bibfield  {journal} {\bibinfo  {journal} {Phys. Rev. X}\ }\textbf {\bibinfo
  {volume} {8}},\ \bibinfo {pages} {031057} (\bibinfo {year}
  {2018})}\BibitemShut {NoStop}%
\bibitem [{\citenamefont {Arute}\ \emph {et~al.}(2019)\citenamefont {Arute},
  \citenamefont {Arya}, \citenamefont {Babbush}, \citenamefont {Bacon},
  \citenamefont {Bardin}, \citenamefont {Barends}, \citenamefont {Biswas},
  \citenamefont {Boixo} \emph {et~al.}}]{arute_quantum_2019}%
  \BibitemOpen
  \bibfield  {author} {\bibinfo {author} {\bibfnamefont {F.}~\bibnamefont
  {Arute}}, \bibinfo {author} {\bibfnamefont {K.}~\bibnamefont {Arya}},
  \bibinfo {author} {\bibfnamefont {R.}~\bibnamefont {Babbush}}, \bibinfo
  {author} {\bibfnamefont {D.}~\bibnamefont {Bacon}}, \bibinfo {author}
  {\bibfnamefont {J.~C.}\ \bibnamefont {Bardin}}, \bibinfo {author}
  {\bibfnamefont {R.}~\bibnamefont {Barends}}, \bibinfo {author} {\bibfnamefont
  {R.}~\bibnamefont {Biswas}}, \bibinfo {author} {\bibfnamefont
  {S.}~\bibnamefont {Boixo}}, \emph {et~al.},\ }\bibfield  {title} {\bibinfo
  {title} {Quantum supremacy using a programmable superconducting processor},\
  }\href {https://doi.org/10.1038/s41586-019-1666-5} {\bibfield  {journal}
  {\bibinfo  {journal} {Nature}\ }\textbf {\bibinfo {volume} {574}},\ \bibinfo
  {pages} {505} (\bibinfo {year} {2019})}\BibitemShut {NoStop}%
\bibitem [{\citenamefont {Zhou}\ and\ \citenamefont
  {Nahum}(2019)}]{zhou_emergent_2019}%
  \BibitemOpen
  \bibfield  {author} {\bibinfo {author} {\bibfnamefont {T.}~\bibnamefont
  {Zhou}}\ and\ \bibinfo {author} {\bibfnamefont {A.}~\bibnamefont {Nahum}},\
  }\bibfield  {title} {\bibinfo {title} {Emergent statistical mechanics of
  entanglement in random unitary circuits},\ }\href
  {https://doi.org/10.1103/PhysRevB.99.174205} {\bibfield  {journal} {\bibinfo
  {journal} {Phys. Rev. B}\ }\textbf {\bibinfo {volume} {99}},\ \bibinfo
  {pages} {174205} (\bibinfo {year} {2019})}\BibitemShut {NoStop}%
\bibitem [{\citenamefont {Piroli}\ \emph {et~al.}(2020)\citenamefont {Piroli},
  \citenamefont {Sunderhauf},\ and\ \citenamefont {Qi}}]{piroli_random_2020}%
  \BibitemOpen
  \bibfield  {author} {\bibinfo {author} {\bibfnamefont {L.}~\bibnamefont
  {Piroli}}, \bibinfo {author} {\bibfnamefont {C.}~\bibnamefont {Sunderhauf}},\
  and\ \bibinfo {author} {\bibfnamefont {X.-L.}\ \bibnamefont {Qi}},\
  }\bibfield  {title} {\bibinfo {title} {A {Random} {Unitary} {Circuit} {Model}
  for {Black} {Hole} {Evaporation}},\ }\href
  {https://doi.org/10.1007/JHEP04(2020)063} {\bibfield  {journal} {\bibinfo
  {journal} {Journal of High Energy Physics}\ }\textbf {\bibinfo {volume}
  {2020}},\ \bibinfo {pages} {63} (\bibinfo {year} {2020})}\BibitemShut
  {NoStop}%
\bibitem [{\citenamefont {Mi}\ \emph {et~al.}(2021)\citenamefont {Mi},
  \citenamefont {Roushan}, \citenamefont {Quintana}, \citenamefont {Mandra},
  \citenamefont {Marshall}, \citenamefont {Neill}, \citenamefont {Arute},
  \citenamefont {Arya} \emph {et~al.}}]{mi_information_2021}%
  \BibitemOpen
  \bibfield  {author} {\bibinfo {author} {\bibfnamefont {X.}~\bibnamefont
  {Mi}}, \bibinfo {author} {\bibfnamefont {P.}~\bibnamefont {Roushan}},
  \bibinfo {author} {\bibfnamefont {C.}~\bibnamefont {Quintana}}, \bibinfo
  {author} {\bibfnamefont {S.}~\bibnamefont {Mandra}}, \bibinfo {author}
  {\bibfnamefont {J.}~\bibnamefont {Marshall}}, \bibinfo {author}
  {\bibfnamefont {C.}~\bibnamefont {Neill}}, \bibinfo {author} {\bibfnamefont
  {F.}~\bibnamefont {Arute}}, \bibinfo {author} {\bibfnamefont
  {K.}~\bibnamefont {Arya}}, \emph {et~al.},\ }\bibfield  {title} {\bibinfo
  {title} {Information scrambling in quantum circuits},\ }\href
  {https://doi.org/10.1126/science.abg5029} {\bibfield  {journal} {\bibinfo
  {journal} {Science}\ }\textbf {\bibinfo {volume} {374}},\ \bibinfo {pages}
  {1479} (\bibinfo {year} {2021})}\BibitemShut {NoStop}%
\bibitem [{\citenamefont {Potter}\ and\ \citenamefont
  {Vasseur}(2022)}]{potter_entanglement_2022}%
  \BibitemOpen
  \bibfield  {author} {\bibinfo {author} {\bibfnamefont {A.~C.}\ \bibnamefont
  {Potter}}\ and\ \bibinfo {author} {\bibfnamefont {R.}~\bibnamefont
  {Vasseur}},\ }\bibfield  {title} {\bibinfo {title} {Entanglement {Dynamics}
  in {Hybrid} {Quantum} {Circuits}},\ }in\ \href
  {https://doi.org/10.1007/978-3-031-03998-0_9} {\emph {\bibinfo {booktitle}
  {Entanglement in {Spin} {Chains}: {From} {Theory} to {Quantum} {Technology}
  {Applications}}}},\ \bibinfo {series and number} {Quantum {Science} and
  {Technology}},\ \bibinfo {editor} {edited by\ \bibinfo {editor}
  {\bibfnamefont {A.}~\bibnamefont {Bayat}}, \bibinfo {editor} {\bibfnamefont
  {S.}~\bibnamefont {Bose}},\ and\ \bibinfo {editor} {\bibfnamefont
  {H.}~\bibnamefont {Johannesson}}}\ (\bibinfo {address} {Cham},\ \bibinfo
  {year} {2022})\ pp.\ \bibinfo {pages} {211--249}\BibitemShut {NoStop}%
\bibitem [{\citenamefont {Fisher}\ \emph {et~al.}(2022)\citenamefont {Fisher},
  \citenamefont {Khemani}, \citenamefont {Nahum},\ and\ \citenamefont
  {Vijay}}]{fisher_random_2022}%
  \BibitemOpen
  \bibfield  {author} {\bibinfo {author} {\bibfnamefont {M.~P.~A.}\
  \bibnamefont {Fisher}}, \bibinfo {author} {\bibfnamefont {V.}~\bibnamefont
  {Khemani}}, \bibinfo {author} {\bibfnamefont {A.}~\bibnamefont {Nahum}},\
  and\ \bibinfo {author} {\bibfnamefont {S.}~\bibnamefont {Vijay}},\ }\bibfield
   {title} {\bibinfo {title} {Random {Quantum} {Circuits}},\ }\bibfield
  {journal} {\bibinfo  {journal} {arXiv e-prints}\ }\href
  {https://doi.org/10.48550/arXiv.2207.14280} {10.48550/arXiv.2207.14280}
  (\bibinfo {year} {2022})\BibitemShut {NoStop}%
\bibitem [{\citenamefont {Skinner}\ \emph {et~al.}(2019)\citenamefont
  {Skinner}, \citenamefont {Ruhman},\ and\ \citenamefont
  {Nahum}}]{skinner_measurement-induced_2019}%
  \BibitemOpen
  \bibfield  {author} {\bibinfo {author} {\bibfnamefont {B.}~\bibnamefont
  {Skinner}}, \bibinfo {author} {\bibfnamefont {J.}~\bibnamefont {Ruhman}},\
  and\ \bibinfo {author} {\bibfnamefont {A.}~\bibnamefont {Nahum}},\ }\bibfield
   {title} {\bibinfo {title} {Measurement-{Induced} {Phase} {Transitions} in
  the {Dynamics} of {Entanglement}},\ }\href
  {https://doi.org/10.1103/PhysRevX.9.031009} {\bibfield  {journal} {\bibinfo
  {journal} {Physical Review X}\ }\textbf {\bibinfo {volume} {9}},\ \bibinfo
  {pages} {031009} (\bibinfo {year} {2019})}\BibitemShut {NoStop}%
\bibitem [{\citenamefont {Li}\ \emph {et~al.}(2018)\citenamefont {Li},
  \citenamefont {Chen},\ and\ \citenamefont {Fisher}}]{li_quantum_2018}%
  \BibitemOpen
  \bibfield  {author} {\bibinfo {author} {\bibfnamefont {Y.}~\bibnamefont
  {Li}}, \bibinfo {author} {\bibfnamefont {X.}~\bibnamefont {Chen}},\ and\
  \bibinfo {author} {\bibfnamefont {M.~P.~A.}\ \bibnamefont {Fisher}},\
  }\bibfield  {title} {\bibinfo {title} {Quantum {Zeno} effect and the
  many-body entanglement transition},\ }\href
  {https://doi.org/10.1103/PhysRevB.98.205136} {\bibfield  {journal} {\bibinfo
  {journal} {Physical Review B}\ }\textbf {\bibinfo {volume} {98}},\ \bibinfo
  {pages} {205136} (\bibinfo {year} {2018})}\BibitemShut {NoStop}%
\bibitem [{\citenamefont {Li}\ \emph {et~al.}(2019)\citenamefont {Li},
  \citenamefont {Chen},\ and\ \citenamefont
  {Fisher}}]{li_measurement-driven_2019}%
  \BibitemOpen
  \bibfield  {author} {\bibinfo {author} {\bibfnamefont {Y.}~\bibnamefont
  {Li}}, \bibinfo {author} {\bibfnamefont {X.}~\bibnamefont {Chen}},\ and\
  \bibinfo {author} {\bibfnamefont {M.~P.~A.}\ \bibnamefont {Fisher}},\
  }\bibfield  {title} {\bibinfo {title} {Measurement-driven entanglement
  transition in hybrid quantum circuits},\ }\href
  {https://doi.org/10.1103/PhysRevB.100.134306} {\bibfield  {journal} {\bibinfo
   {journal} {Physical Review B}\ }\textbf {\bibinfo {volume} {100}},\ \bibinfo
  {pages} {134306} (\bibinfo {year} {2019})}\BibitemShut {NoStop}%
\bibitem [{\citenamefont {Gullans}\ and\ \citenamefont
  {Huse}(2020{\natexlab{a}})}]{gullans_dynamical_2020}%
  \BibitemOpen
  \bibfield  {author} {\bibinfo {author} {\bibfnamefont {M.~J.}\ \bibnamefont
  {Gullans}}\ and\ \bibinfo {author} {\bibfnamefont {D.~A.}\ \bibnamefont
  {Huse}},\ }\bibfield  {title} {\bibinfo {title} {Dynamical {Purification}
  {Phase} {Transition} {Induced} by {Quantum} {Measurements}},\ }\href
  {https://doi.org/10.1103/PhysRevX.10.041020} {\bibfield  {journal} {\bibinfo
  {journal} {Physical Review X}\ }\textbf {\bibinfo {volume} {10}},\ \bibinfo
  {pages} {041020} (\bibinfo {year} {2020}{\natexlab{a}})}\BibitemShut
  {NoStop}%
\bibitem [{\citenamefont {Choi}\ \emph {et~al.}(2020)\citenamefont {Choi},
  \citenamefont {Bao}, \citenamefont {Qi},\ and\ \citenamefont
  {Altman}}]{choi_quantum_2020}%
  \BibitemOpen
  \bibfield  {author} {\bibinfo {author} {\bibfnamefont {S.}~\bibnamefont
  {Choi}}, \bibinfo {author} {\bibfnamefont {Y.}~\bibnamefont {Bao}}, \bibinfo
  {author} {\bibfnamefont {X.-L.}\ \bibnamefont {Qi}},\ and\ \bibinfo {author}
  {\bibfnamefont {E.}~\bibnamefont {Altman}},\ }\bibfield  {title} {\bibinfo
  {title} {Quantum {Error} {Correction} in {Scrambling} {Dynamics} and
  {Measurement}-{Induced} {Phase} {Transition}},\ }\href
  {https://doi.org/10.1103/PhysRevLett.125.030505} {\bibfield  {journal}
  {\bibinfo  {journal} {Physical Review Letters}\ }\textbf {\bibinfo {volume}
  {125}},\ \bibinfo {pages} {030505} (\bibinfo {year} {2020})}\BibitemShut
  {NoStop}%
\bibitem [{\citenamefont {Eisert}\ \emph {et~al.}(2010)\citenamefont {Eisert},
  \citenamefont {Cramer},\ and\ \citenamefont {Plenio}}]{eisert_arealaw_2010}%
  \BibitemOpen
  \bibfield  {author} {\bibinfo {author} {\bibfnamefont {J.}~\bibnamefont
  {Eisert}}, \bibinfo {author} {\bibfnamefont {M.}~\bibnamefont {Cramer}},\
  and\ \bibinfo {author} {\bibfnamefont {M.~B.}\ \bibnamefont {Plenio}},\
  }\bibfield  {title} {\bibinfo {title} {Colloquium: Area laws for the
  entanglement entropy},\ }\href {https://doi.org/10.1103/RevModPhys.82.277}
  {\bibfield  {journal} {\bibinfo  {journal} {Rev. Mod. Phys.}\ }\textbf
  {\bibinfo {volume} {82}},\ \bibinfo {pages} {277} (\bibinfo {year}
  {2010})}\BibitemShut {NoStop}%
\bibitem [{\citenamefont {Page}(1993)}]{page_entropy_1993}%
  \BibitemOpen
  \bibfield  {author} {\bibinfo {author} {\bibfnamefont {D.~N.}\ \bibnamefont
  {Page}},\ }\bibfield  {title} {\bibinfo {title} {Average entropy of a
  subsystem},\ }\href {https://doi.org/10.1103/PhysRevLett.71.1291} {\bibfield
  {journal} {\bibinfo  {journal} {Phys. Rev. Lett.}\ }\textbf {\bibinfo
  {volume} {71}},\ \bibinfo {pages} {1291} (\bibinfo {year}
  {1993})}\BibitemShut {NoStop}%
\bibitem [{\citenamefont {Jian}\ \emph {et~al.}(2020)\citenamefont {Jian},
  \citenamefont {You}, \citenamefont {Vasseur},\ and\ \citenamefont
  {Ludwig}}]{jian_measurement-induced_2020}%
  \BibitemOpen
  \bibfield  {author} {\bibinfo {author} {\bibfnamefont {C.-M.}\ \bibnamefont
  {Jian}}, \bibinfo {author} {\bibfnamefont {Y.-Z.}\ \bibnamefont {You}},
  \bibinfo {author} {\bibfnamefont {R.}~\bibnamefont {Vasseur}},\ and\ \bibinfo
  {author} {\bibfnamefont {A.~W.~W.}\ \bibnamefont {Ludwig}},\ }\bibfield
  {title} {\bibinfo {title} {Measurement-induced criticality in random quantum
  circuits},\ }\href {https://doi.org/10.1103/PhysRevB.101.104302} {\bibfield
  {journal} {\bibinfo  {journal} {Physical Review B}\ }\textbf {\bibinfo
  {volume} {101}},\ \bibinfo {pages} {104302} (\bibinfo {year}
  {2020})}\BibitemShut {NoStop}%
\bibitem [{\citenamefont {Bao}\ \emph {et~al.}(2020)\citenamefont {Bao},
  \citenamefont {Choi},\ and\ \citenamefont {Altman}}]{bao_theory_2020}%
  \BibitemOpen
  \bibfield  {author} {\bibinfo {author} {\bibfnamefont {Y.}~\bibnamefont
  {Bao}}, \bibinfo {author} {\bibfnamefont {S.}~\bibnamefont {Choi}},\ and\
  \bibinfo {author} {\bibfnamefont {E.}~\bibnamefont {Altman}},\ }\bibfield
  {title} {\bibinfo {title} {Theory of the phase transition in random unitary
  circuits with measurements},\ }\href
  {https://doi.org/10.1103/PhysRevB.101.104301} {\bibfield  {journal} {\bibinfo
   {journal} {Physical Review B}\ }\textbf {\bibinfo {volume} {101}},\ \bibinfo
  {pages} {104301} (\bibinfo {year} {2020})}\BibitemShut {NoStop}%
\bibitem [{\citenamefont {Li}\ \emph {et~al.}(2021)\citenamefont {Li},
  \citenamefont {Chen}, \citenamefont {Ludwig},\ and\ \citenamefont
  {Fisher}}]{li_conformal_2021}%
  \BibitemOpen
  \bibfield  {author} {\bibinfo {author} {\bibfnamefont {Y.}~\bibnamefont
  {Li}}, \bibinfo {author} {\bibfnamefont {X.}~\bibnamefont {Chen}}, \bibinfo
  {author} {\bibfnamefont {A.~W.~W.}\ \bibnamefont {Ludwig}},\ and\ \bibinfo
  {author} {\bibfnamefont {M.~P.~A.}\ \bibnamefont {Fisher}},\ }\bibfield
  {title} {\bibinfo {title} {Conformal invariance and quantum nonlocality in
  critical hybrid circuits},\ }\href
  {https://doi.org/10.1103/PhysRevB.104.104305} {\bibfield  {journal} {\bibinfo
   {journal} {Physical Review B}\ }\textbf {\bibinfo {volume} {104}},\ \bibinfo
  {pages} {104305} (\bibinfo {year} {2021})}\BibitemShut {NoStop}%
\bibitem [{\citenamefont {Jonay}\ \emph {et~al.}(2018)\citenamefont {Jonay},
  \citenamefont {Huse},\ and\ \citenamefont
  {Nahum}}]{jonay_coarse-grained_2018}%
  \BibitemOpen
  \bibfield  {author} {\bibinfo {author} {\bibfnamefont {C.}~\bibnamefont
  {Jonay}}, \bibinfo {author} {\bibfnamefont {D.~A.}\ \bibnamefont {Huse}},\
  and\ \bibinfo {author} {\bibfnamefont {A.}~\bibnamefont {Nahum}},\ }\bibfield
   {title} {\bibinfo {title} {Coarse-grained dynamics of operator and state
  entanglement}\ }\href {https://doi.org/10.48550/arXiv.1803.00089}
  {10.48550/arXiv.1803.00089} (\bibinfo {year} {2018})\BibitemShut {NoStop}%
\bibitem [{\citenamefont {Li}\ and\ \citenamefont
  {Fisher}(2021)}]{li_statistical_2021}%
  \BibitemOpen
  \bibfield  {author} {\bibinfo {author} {\bibfnamefont {Y.}~\bibnamefont
  {Li}}\ and\ \bibinfo {author} {\bibfnamefont {M.~P.~A.}\ \bibnamefont
  {Fisher}},\ }\bibfield  {title} {\bibinfo {title} {Statistical mechanics of
  quantum error correcting codes},\ }\href
  {https://doi.org/10.1103/PhysRevB.103.104306} {\bibfield  {journal} {\bibinfo
   {journal} {Physical Review B}\ }\textbf {\bibinfo {volume} {103}},\ \bibinfo
  {pages} {104306} (\bibinfo {year} {2021})}\BibitemShut {NoStop}%
\bibitem [{\citenamefont {Li}\ \emph {et~al.}(2023)\citenamefont {Li},
  \citenamefont {Vijay},\ and\ \citenamefont {Fisher}}]{li_entanglement_2023}%
  \BibitemOpen
  \bibfield  {author} {\bibinfo {author} {\bibfnamefont {Y.}~\bibnamefont
  {Li}}, \bibinfo {author} {\bibfnamefont {S.}~\bibnamefont {Vijay}},\ and\
  \bibinfo {author} {\bibfnamefont {M.~P.}\ \bibnamefont {Fisher}},\ }\bibfield
   {title} {\bibinfo {title} {Entanglement {Domain} {Walls} in {Monitored}
  {Quantum} {Circuits} and the {Directed} {Polymer} in a {Random}
  {Environment}},\ }\href {https://doi.org/10.1103/PRXQuantum.4.010331}
  {\bibfield  {journal} {\bibinfo  {journal} {PRX Quantum}\ }\textbf {\bibinfo
  {volume} {4}},\ \bibinfo {pages} {010331} (\bibinfo {year}
  {2023})}\BibitemShut {NoStop}%
\bibitem [{\citenamefont {Pastawski}\ \emph {et~al.}(2015)\citenamefont
  {Pastawski}, \citenamefont {Yoshida}, \citenamefont {Harlow},\ and\
  \citenamefont {Preskill}}]{pastawski_holographic_2015}%
  \BibitemOpen
  \bibfield  {author} {\bibinfo {author} {\bibfnamefont {F.}~\bibnamefont
  {Pastawski}}, \bibinfo {author} {\bibfnamefont {B.}~\bibnamefont {Yoshida}},
  \bibinfo {author} {\bibfnamefont {D.}~\bibnamefont {Harlow}},\ and\ \bibinfo
  {author} {\bibfnamefont {J.}~\bibnamefont {Preskill}},\ }\bibfield  {title}
  {\bibinfo {title} {Holographic quantum error-correcting codes: toy models for
  the bulk/boundary correspondence},\ }\href
  {https://doi.org/10.1007/JHEP06(2015)149} {\bibfield  {journal} {\bibinfo
  {journal} {Journal of High Energy Physics}\ }\textbf {\bibinfo {volume}
  {2015}},\ \bibinfo {pages} {149} (\bibinfo {year} {2015})}\BibitemShut
  {NoStop}%
\bibitem [{\citenamefont {Hayden}\ \emph {et~al.}(2016)\citenamefont {Hayden},
  \citenamefont {Nezami}, \citenamefont {Qi}, \citenamefont {Thomas},
  \citenamefont {Walter},\ and\ \citenamefont {Yang}}]{hayden2016holographic}%
  \BibitemOpen
  \bibfield  {author} {\bibinfo {author} {\bibfnamefont {P.}~\bibnamefont
  {Hayden}}, \bibinfo {author} {\bibfnamefont {S.}~\bibnamefont {Nezami}},
  \bibinfo {author} {\bibfnamefont {X.-L.}\ \bibnamefont {Qi}}, \bibinfo
  {author} {\bibfnamefont {N.}~\bibnamefont {Thomas}}, \bibinfo {author}
  {\bibfnamefont {M.}~\bibnamefont {Walter}},\ and\ \bibinfo {author}
  {\bibfnamefont {Z.}~\bibnamefont {Yang}},\ }\bibfield  {title} {\bibinfo
  {title} {Holographic duality from random tensor networks},\ }\href@noop {}
  {\bibfield  {journal} {\bibinfo  {journal} {Journal of High Energy Physics}\
  }\textbf {\bibinfo {volume} {2016}},\ \bibinfo {pages} {1} (\bibinfo {year}
  {2016})}\BibitemShut {NoStop}%
\bibitem [{\citenamefont {Jahn}\ and\ \citenamefont
  {Eisert}(2021)}]{jahn2021holographic}%
  \BibitemOpen
  \bibfield  {author} {\bibinfo {author} {\bibfnamefont {A.}~\bibnamefont
  {Jahn}}\ and\ \bibinfo {author} {\bibfnamefont {J.}~\bibnamefont {Eisert}},\
  }\bibfield  {title} {\bibinfo {title} {Holographic tensor network models and
  quantum error correction: a topical review},\ }\href@noop {} {\bibfield
  {journal} {\bibinfo  {journal} {Quantum Science and Technology}\ }\textbf
  {\bibinfo {volume} {6}},\ \bibinfo {pages} {033002} (\bibinfo {year}
  {2021})}\BibitemShut {NoStop}%
\bibitem [{\citenamefont {Banados}\ \emph {et~al.}(1992)\citenamefont
  {Banados}, \citenamefont {Teitelboim},\ and\ \citenamefont
  {Zanelli}}]{banados_black_1992}%
  \BibitemOpen
  \bibfield  {author} {\bibinfo {author} {\bibfnamefont {M.}~\bibnamefont
  {Banados}}, \bibinfo {author} {\bibfnamefont {C.}~\bibnamefont
  {Teitelboim}},\ and\ \bibinfo {author} {\bibfnamefont {J.}~\bibnamefont
  {Zanelli}},\ }\bibfield  {title} {\bibinfo {title} {Black hole in
  three-dimensional spacetime},\ }\href
  {https://doi.org/10.1103/PhysRevLett.69.1849} {\bibfield  {journal} {\bibinfo
   {journal} {Physical Review Letters}\ }\textbf {\bibinfo {volume} {69}},\
  \bibinfo {pages} {1849} (\bibinfo {year} {1992})}\BibitemShut {NoStop}%
\bibitem [{\citenamefont {Aaronson}\ and\ \citenamefont
  {Gottesman}(2004)}]{aaronson_improved_2004}%
  \BibitemOpen
  \bibfield  {author} {\bibinfo {author} {\bibfnamefont {S.}~\bibnamefont
  {Aaronson}}\ and\ \bibinfo {author} {\bibfnamefont {D.}~\bibnamefont
  {Gottesman}},\ }\bibfield  {title} {\bibinfo {title} {Improved simulation of
  stabilizer circuits},\ }\href {https://doi.org/10.1103/PhysRevA.70.052328}
  {\bibfield  {journal} {\bibinfo  {journal} {Phys. Rev. A}\ }\textbf {\bibinfo
  {volume} {70}},\ \bibinfo {pages} {052328} (\bibinfo {year}
  {2004})}\BibitemShut {NoStop}%
\bibitem [{\citenamefont {Vasseur}\ \emph {et~al.}(2019)\citenamefont
  {Vasseur}, \citenamefont {Potter}, \citenamefont {You},\ and\ \citenamefont
  {Ludwig}}]{vasseur_entanglement_2019}%
  \BibitemOpen
  \bibfield  {author} {\bibinfo {author} {\bibfnamefont {R.}~\bibnamefont
  {Vasseur}}, \bibinfo {author} {\bibfnamefont {A.~C.}\ \bibnamefont {Potter}},
  \bibinfo {author} {\bibfnamefont {Y.-Z.}\ \bibnamefont {You}},\ and\ \bibinfo
  {author} {\bibfnamefont {A.~W.~W.}\ \bibnamefont {Ludwig}},\ }\bibfield
  {title} {\bibinfo {title} {Entanglement transitions from holographic random
  tensor networks},\ }\href {https://doi.org/10.1103/PhysRevB.100.134203}
  {\bibfield  {journal} {\bibinfo  {journal} {Physical Review B}\ }\textbf
  {\bibinfo {volume} {100}},\ \bibinfo {pages} {134203} (\bibinfo {year}
  {2019})}\BibitemShut {NoStop}%
\bibitem [{\citenamefont {Zabalo}\ \emph {et~al.}(2020)\citenamefont {Zabalo},
  \citenamefont {Gullans}, \citenamefont {Wilson}, \citenamefont
  {Gopalakrishnan}, \citenamefont {Huse},\ and\ \citenamefont
  {Pixley}}]{zabalo_critical_2020}%
  \BibitemOpen
  \bibfield  {author} {\bibinfo {author} {\bibfnamefont {A.}~\bibnamefont
  {Zabalo}}, \bibinfo {author} {\bibfnamefont {M.~J.}\ \bibnamefont {Gullans}},
  \bibinfo {author} {\bibfnamefont {J.~H.}\ \bibnamefont {Wilson}}, \bibinfo
  {author} {\bibfnamefont {S.}~\bibnamefont {Gopalakrishnan}}, \bibinfo
  {author} {\bibfnamefont {D.~A.}\ \bibnamefont {Huse}},\ and\ \bibinfo
  {author} {\bibfnamefont {J.~H.}\ \bibnamefont {Pixley}},\ }\bibfield  {title}
  {\bibinfo {title} {Critical properties of the measurement-induced transition
  in random quantum circuits},\ }\href
  {https://doi.org/10.1103/PhysRevB.101.060301} {\bibfield  {journal} {\bibinfo
   {journal} {Physical Review B}\ }\textbf {\bibinfo {volume} {101}},\ \bibinfo
  {pages} {060301} (\bibinfo {year} {2020})}\BibitemShut {NoStop}%
\bibitem [{Note1()}]{Note1}%
  \BibitemOpen
  \bibinfo {note} {In the sense that, for a uniform metric, one has $S_A \sim
  s|A|$ for an interval in a late-time state (geodesic directed along $\protect
  \hat {x}$) and $S \sim sv_E \Delta t$ for a half-cut in a product state
  evolved for a short time $\Delta t$ (geodesic directed along $\protect \hat
  {t}$).}\BibitemShut {Stop}%
\bibitem [{\citenamefont {Bertini}\ \emph {et~al.}(2019)\citenamefont
  {Bertini}, \citenamefont {Kos},\ and\ \citenamefont
  {Prosen}}]{bertini_exact_2019}%
  \BibitemOpen
  \bibfield  {author} {\bibinfo {author} {\bibfnamefont {B.}~\bibnamefont
  {Bertini}}, \bibinfo {author} {\bibfnamefont {P.}~\bibnamefont {Kos}},\ and\
  \bibinfo {author} {\bibfnamefont {T.}~\bibnamefont {Prosen}},\ }\bibfield
  {title} {\bibinfo {title} {Exact {Correlation} {Functions} for
  {Dual}-{Unitary} {Lattice} {Models} in \$1+1\$ {Dimensions}},\ }\href
  {https://doi.org/10.1103/PhysRevLett.123.210601} {\bibfield  {journal}
  {\bibinfo  {journal} {Phys. Rev. Lett.}\ }\textbf {\bibinfo {volume} {123}},\
  \bibinfo {pages} {210601} (\bibinfo {year} {2019})}\BibitemShut {NoStop}%
\bibitem [{Note2()}]{Note2}%
  \BibitemOpen
  \bibinfo {note} {The fact that $v_E=1$ in these circuits was previously
  exploited to simplify the extraction of critical exponents at the
  transition~\cite {zabalo_operator_2022}.}\BibitemShut {Stop}%
\bibitem [{\citenamefont {Zabalo}\ \emph
  {et~al.}(2022{\natexlab{a}})\citenamefont {Zabalo}, \citenamefont {Gullans},
  \citenamefont {Wilson}, \citenamefont {Vasseur}, \citenamefont {Ludwig},
  \citenamefont {Gopalakrishnan}, \citenamefont {Huse},\ and\ \citenamefont
  {Pixley}}]{zabalo_operator_2022}%
  \BibitemOpen
  \bibfield  {author} {\bibinfo {author} {\bibfnamefont {A.}~\bibnamefont
  {Zabalo}}, \bibinfo {author} {\bibfnamefont {M.~J.}\ \bibnamefont {Gullans}},
  \bibinfo {author} {\bibfnamefont {J.~H.}\ \bibnamefont {Wilson}}, \bibinfo
  {author} {\bibfnamefont {R.}~\bibnamefont {Vasseur}}, \bibinfo {author}
  {\bibfnamefont {A.~W.~W.}\ \bibnamefont {Ludwig}}, \bibinfo {author}
  {\bibfnamefont {S.}~\bibnamefont {Gopalakrishnan}}, \bibinfo {author}
  {\bibfnamefont {D.~A.}\ \bibnamefont {Huse}},\ and\ \bibinfo {author}
  {\bibfnamefont {J.~H.}\ \bibnamefont {Pixley}},\ }\bibfield  {title}
  {\bibinfo {title} {Operator {Scaling} {Dimensions} and {Multifractality} at
  {Measurement}-{Induced} {Transitions}},\ }\href
  {https://doi.org/10.1103/PhysRevLett.128.050602} {\bibfield  {journal}
  {\bibinfo  {journal} {Physical Review Letters}\ }\textbf {\bibinfo {volume}
  {128}},\ \bibinfo {pages} {050602} (\bibinfo {year}
  {2022}{\natexlab{a}})}\BibitemShut {NoStop}%
\bibitem [{Note3()}]{Note3}%
  \BibitemOpen
  \bibinfo {note} {See online Supplementary Material for numerical study of the
  entanglement transition, derivation of the BTZ measurement rate and geodesic
  length, and numerics on AdS states.}\BibitemShut {Stop}%
\bibitem [{\citenamefont {Bekenstein}(1981)}]{bekenstein1981universal}%
  \BibitemOpen
  \bibfield  {author} {\bibinfo {author} {\bibfnamefont {J.~D.}\ \bibnamefont
  {Bekenstein}},\ }\bibfield  {title} {\bibinfo {title} {Universal upper bound
  on the entropy-to-energy ratio for bounded systems},\ }\href@noop {}
  {\bibfield  {journal} {\bibinfo  {journal} {Physical Review D}\ }\textbf
  {\bibinfo {volume} {23}},\ \bibinfo {pages} {287} (\bibinfo {year}
  {1981})}\BibitemShut {NoStop}%
\bibitem [{\citenamefont {Headrick}(2010)}]{headrick2010entanglement}%
  \BibitemOpen
  \bibfield  {author} {\bibinfo {author} {\bibfnamefont {M.}~\bibnamefont
  {Headrick}},\ }\bibfield  {title} {\bibinfo {title} {Entanglement renyi
  entropies in holographic theories},\ }\href@noop {} {\bibfield  {journal}
  {\bibinfo  {journal} {Physical Review D}\ }\textbf {\bibinfo {volume} {82}},\
  \bibinfo {pages} {126010} (\bibinfo {year} {2010})}\BibitemShut {NoStop}%
\bibitem [{\citenamefont {Gullans}\ and\ \citenamefont
  {Huse}(2020{\natexlab{b}})}]{gullans_scalable_2020}%
  \BibitemOpen
  \bibfield  {author} {\bibinfo {author} {\bibfnamefont {M.~J.}\ \bibnamefont
  {Gullans}}\ and\ \bibinfo {author} {\bibfnamefont {D.~A.}\ \bibnamefont
  {Huse}},\ }\bibfield  {title} {\bibinfo {title} {Scalable {Probes} of
  {Measurement}-{Induced} {Criticality}},\ }\href
  {https://doi.org/10.1103/PhysRevLett.125.070606} {\bibfield  {journal}
  {\bibinfo  {journal} {Physical Review Letters}\ }\textbf {\bibinfo {volume}
  {125}},\ \bibinfo {pages} {070606} (\bibinfo {year}
  {2020}{\natexlab{b}})}\BibitemShut {NoStop}%
\bibitem [{\citenamefont {Ippoliti}\ \emph {et~al.}(2021)\citenamefont
  {Ippoliti}, \citenamefont {Gullans}, \citenamefont {Gopalakrishnan},
  \citenamefont {Huse},\ and\ \citenamefont
  {Khemani}}]{ippoliti_entanglement_2021}%
  \BibitemOpen
  \bibfield  {author} {\bibinfo {author} {\bibfnamefont {M.}~\bibnamefont
  {Ippoliti}}, \bibinfo {author} {\bibfnamefont {M.~J.}\ \bibnamefont
  {Gullans}}, \bibinfo {author} {\bibfnamefont {S.}~\bibnamefont
  {Gopalakrishnan}}, \bibinfo {author} {\bibfnamefont {D.~A.}\ \bibnamefont
  {Huse}},\ and\ \bibinfo {author} {\bibfnamefont {V.}~\bibnamefont
  {Khemani}},\ }\bibfield  {title} {\bibinfo {title} {Entanglement {Phase}
  {Transitions} in {Measurement}-{Only} {Dynamics}},\ }\href
  {https://doi.org/10.1103/PhysRevX.11.011030} {\bibfield  {journal} {\bibinfo
  {journal} {Physical Review X}\ }\textbf {\bibinfo {volume} {11}},\ \bibinfo
  {pages} {011030} (\bibinfo {year} {2021})}\BibitemShut {NoStop}%
\bibitem [{Note4()}]{Note4}%
  \BibitemOpen
  \bibinfo {note} {As a practical matter in simulations, after performing a
  measurement on a site, we may entangle the qubit at that location with a
  reference $R$ instead of leaving it in the measured state.}\BibitemShut
  {Stop}%
\bibitem [{Note5()}]{Note5}%
  \BibitemOpen
  \bibinfo {note} {It is always possible that the location where $R$ is
  entangled gets measured again immediately afterwards, disentangling $R$ with
  some probability even inside the wedge.}\BibitemShut {Stop}%
\bibitem [{\citenamefont {Zabalo}\ \emph
  {et~al.}(2022{\natexlab{b}})\citenamefont {Zabalo}, \citenamefont {Wilson},
  \citenamefont {Gullans}, \citenamefont {Vasseur}, \citenamefont
  {Gopalakrishnan}, \citenamefont {Huse},\ and\ \citenamefont
  {Pixley}}]{zabalo_infinite-randomness_2022}%
  \BibitemOpen
  \bibfield  {author} {\bibinfo {author} {\bibfnamefont {A.}~\bibnamefont
  {Zabalo}}, \bibinfo {author} {\bibfnamefont {J.~H.}\ \bibnamefont {Wilson}},
  \bibinfo {author} {\bibfnamefont {M.~J.}\ \bibnamefont {Gullans}}, \bibinfo
  {author} {\bibfnamefont {R.}~\bibnamefont {Vasseur}}, \bibinfo {author}
  {\bibfnamefont {S.}~\bibnamefont {Gopalakrishnan}}, \bibinfo {author}
  {\bibfnamefont {D.~A.}\ \bibnamefont {Huse}},\ and\ \bibinfo {author}
  {\bibfnamefont {J.~H.}\ \bibnamefont {Pixley}},\ }\bibfield  {title}
  {\bibinfo {title} {Infinite-randomness criticality in monitored quantum
  dynamics with static disorder}\ }\href
  {https://doi.org/10.48550/arXiv.2205.14002} {10.48550/arXiv.2205.14002}
  (\bibinfo {year} {2022}{\natexlab{b}})\BibitemShut {NoStop}%
\bibitem [{\citenamefont {Weinstein}\ \emph {et~al.}(2022)\citenamefont
  {Weinstein}, \citenamefont {Bao},\ and\ \citenamefont
  {Altman}}]{weinstein_measurement-induced_2022}%
  \BibitemOpen
  \bibfield  {author} {\bibinfo {author} {\bibfnamefont {Z.}~\bibnamefont
  {Weinstein}}, \bibinfo {author} {\bibfnamefont {Y.}~\bibnamefont {Bao}},\
  and\ \bibinfo {author} {\bibfnamefont {E.}~\bibnamefont {Altman}},\
  }\bibfield  {title} {\bibinfo {title} {Measurement-{Induced} {Power}-{Law}
  {Negativity} in an {Open} {Monitored} {Quantum} {Circuit}},\ }\href
  {https://doi.org/10.1103/PhysRevLett.129.080501} {\bibfield  {journal}
  {\bibinfo  {journal} {Physical Review Letters}\ }\textbf {\bibinfo {volume}
  {129}},\ \bibinfo {pages} {080501} (\bibinfo {year} {2022})}\BibitemShut
  {NoStop}%
\bibitem [{\citenamefont {Maxfield}(2015)}]{maxfield2015entanglement}%
  \BibitemOpen
  \bibfield  {author} {\bibinfo {author} {\bibfnamefont {H.}~\bibnamefont
  {Maxfield}},\ }\bibfield  {title} {\bibinfo {title} {Entanglement entropy in
  three dimensional gravity},\ }\href@noop {} {\bibfield  {journal} {\bibinfo
  {journal} {Journal of High Energy Physics}\ }\textbf {\bibinfo {volume}
  {2015}},\ \bibinfo {pages} {1} (\bibinfo {year} {2015})}\BibitemShut
  {NoStop}%
\end{thebibliography}%

\clearpage
\widetext

\setcounter{equation}{0}
\setcounter{figure}{0}
\setcounter{table}{0}
\setcounter{page}{1}
\makeatletter
\renewcommand{\thesection}{S\arabic{section}}
\renewcommand{\theequation}{S\arabic{equation}}
\renewcommand{\thefigure}{S\arabic{figure}}
\renewcommand{\thepage}{S\arabic{page}}

\begin{center}
\textbf{\large Supplemental Material: \titleplaceholder } \\~\\
Aditya Cowsik, Matteo Ippoliti and Xiaoliang Qi \\
\textit{\small Department of Physics, Stanford University, Stanford, CA, 94305}
\end{center}


\section{Entanglement transition \label{app:transition}}

Here we locate the measurement-induced entanglement transition in the class of circuits used in the main text: 1D brickwork circuits of dual-unitary Clifford gates and projective measurements.
This model was already studied in Ref.~\cite{zabalo_operator_2022}; we independently verify this to ensure consistency of the other simulations in our work.

\begin{figure}[h!]
    \centering
    \includegraphics[width=0.8\textwidth]{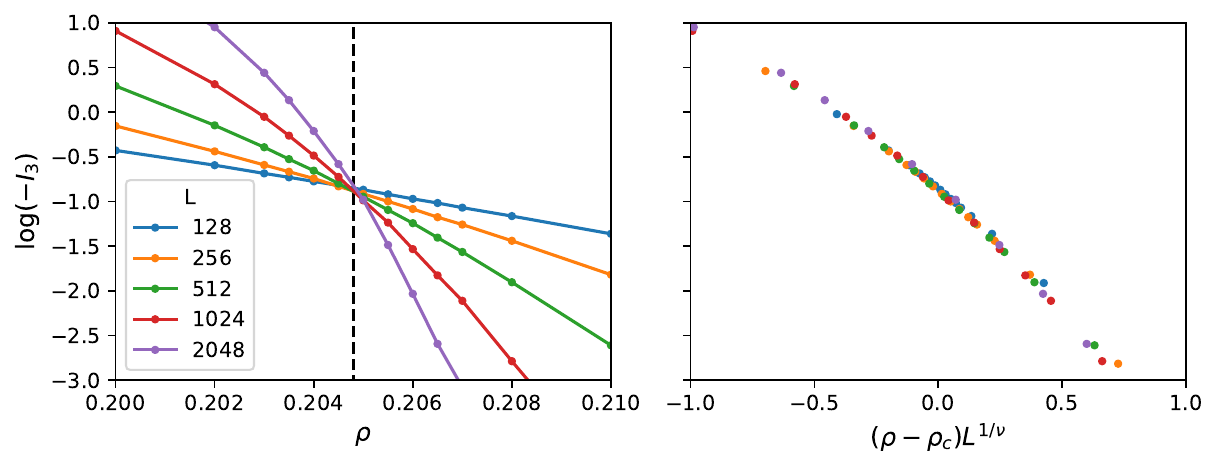}
    \caption{Entanglement phase transition in brickwork circuits of dual-unitary Clifford gates and single-qubit measurements. 
    (a) Tripartite mutual information $I_3(A:B:C)$ between contiguous segments of length $L/4$ as a function of measurement density $\rho$ for different system sizes $L$. Data averaged over between $250$ and $10^4$ realizations (depending on $L$) and over time steps $2L < t < 4L$.
    A finite-size crossing is visible near $\rho = \rho_c \simeq 0.2048$ (vertical dashed line).
    (b) Scaling collapse of the data as a function of $(\rho - \rho_c) L^{1/\nu}$, with correlation length critical exponent $\nu = 1.30$. }
    \label{fig:scaling_collapse}
\end{figure}

To locate the transition, we use the tripartite mutual information $I_3(A:B:C)$ between three regions $A = [0,L/4)$, $B = [L/4,L/2)$, $C=[L/2,3L/4)$ (the system, of length $L$, has periodic boundary conditions). This quantity is known to have limited finite-size drift at the critical point~\cite{zabalo_critical_2020}. 
Results shown in Fig.~\ref{fig:scaling_collapse} indicate a phase transition at $\rho_c = 0.2048(5)$ and critical exponent $\nu = 1.30(5)$, consistent with what is reported in Ref.~\cite{zabalo_operator_2022}. We use these values in the rest of our work.

\section{Derivation of the measurement rate for BTZ metric \label{app:BTZ}}

The metric for a general (rotating) BTZ black hole is~\cite{banados_black_1992}
\begin{equation}
    \diff \ell^2 =  -\frac{(r^2 - r_+^2)(r^2 - r_-^2)}{l^2 r^2}\diff t^2 + \frac{l^2 r^2 \diff r^2}{(r^2 - r_+^2)(r^2 - r_-^2)} + r^2 \left(\diff \phi - \frac{r_+ r_-}{l r^2} \diff t \right)^2,
\end{equation}
where $r_+$ and $r_-$ the radii of the two horizons, and $l$ is AdS radius.
Because the type of metrics we can access are Euclidean and not Lorentzian we choose a constant time slice so that $t = 0$. Also for simplicity we choose a non-rotating black hole, so that $r_{-} = 0$ and
\begin{equation}
    \diff \ell^2 = \frac{l^2 dr^2}{r^2 - r_h^2} + r^2 \diff \phi^2.
\end{equation}

We look for a reparametrization $(x(r,\phi), t(r,\phi))$ such that the metric assumes the isotropic form $d\ell^2 = s(x, t)^2 (dx^2 + dt^2)$ used in the main text. Let us make the ansatz that $\phi(x, t) = \phi(x)$ and $r(x, t) = r(t)$. Then, using the invariance of $\diff \ell^2$ under this reparametrization and varying only $x$ or only $t$ yields
\begin{align}
    \frac{l}{\sqrt{r^2-r_h^2}} \diff r & = s(x, t) \diff t, \\
    r \diff \phi & = s(x, t) \diff x.
\end{align}
We can further set $\phi = 2\pi x/L$ (note that $x \in [0,L)$ while $\phi \in [0,2\pi)$) so that
\begin{align}
    s(x, t) & = \frac{2\pi}{L} r(t), \\
    \frac{Ll}{2\pi r \sqrt{r^2-r_h^2}} \diff r & = \diff t.
\end{align}
Upon integration, the latter yields
\begin{equation}
    t = -T\left( 1-\frac{2}{\pi} \arctan \sqrt{\left(\frac{r}{r_h}\right)^2 - 1}\right)
\end{equation}
where we have set $T = Ll/4r_h$ and chosen the additive constant of integration in such a way that $t \in [-T,0]$, with $r = r_h$ giving $t = -T$ (initial state) and $r\to\infty$ giving $t = 0$ (final state).
Inverting this relation for $r(t)$ yields
\begin{equation}
    r(t) = r_h \left| \sec\left[\frac{\pi}{2}\left(1+\frac{t}{T}\right) \right] \right| = r_h \csc \left( \frac{\pi |t|}{2 T}\right).
\end{equation}

Finally, we obtain the entropy density $s(t)$ reported in the main text:
\begin{equation}
    s(t) = \frac{2\pi}{L} r(t) = \frac{\pi l}{2T} \csc\left( \frac{\pi|t|}{2T} \right).
\end{equation}

Notice that, once $T$ is set, the AdS radius $l$ only controls the scale factor of the metric and thus will not affect the shape of the geodesics. 

\section{Derivation of the Geodesic Length for the BTZ Metric}
Now we would like to compute the distance between two points on the boundary. Because we cut off the evolution before the measurement density, $\rho$, becomes negative, our state is effectively at a finite radius, $r$. In other words our discretization of the space won't allow us to reach infinite radius. Using the result in equation 2.5 for the distance, along with the parameterization in equation 3.2 from \cite{maxfield2015entanglement} at $t = 0$ allows us to compute the distance between two points at location $\phi_1$ and $\phi_2$ and at the same radius $r$. This radius models the actual, finite, radius the state is located at due to the cutoff. A straightforward substitution and calculation yields
\begin{equation}
    d(\phi_1,\phi_2) = \min_{n \in \mathbb{Z}} \cosh^{-1} \left[\frac{r^2}{r_h^2}\left(\cosh\left(r_h (\phi_1-\phi_2 + 2 \pi n) \right) -1 \right) +1 \right],
\end{equation}
where minimizing over all $n \in \mathbb{Z}$ follows from the fact that we want the shortest path, avoiding geodesics which wind around the black hole. We can therefore fit the metric on the boundary with these two parameters along with an additional parameter for the entropy per unit length in the space. Of course there is one additional fact we should take care of, which is that the maximum entropy is limited in our discrete model since we impose an explicit UV cutoff. Therefore the entropy fitting should be given by the minimum of the distance and the number of qubits in the region, since we cannot have more entropy than the number of qubits. 


\section{Logarithmically-entangled states from hyperbolic metric \label{app:AdS}}

\begin{figure}[h!]
    \centering
    \includegraphics[width=\textwidth]{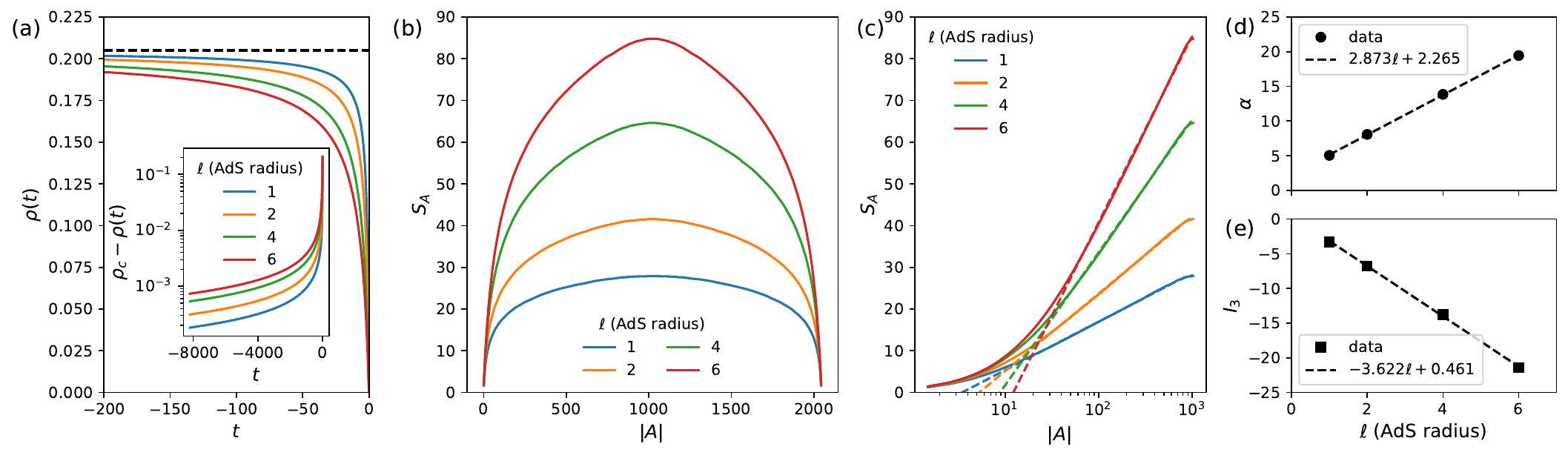}
    \caption{Logarithmically-entangled states from the AdS metric. 
    (a) Measurement density as a function of time: $\rho(t) = \rho_c[1-(1+|t|/l)^{-1/\nu}]$, for different values of the parameter $l$ which plays the role of AdS radius. The dashed horizontal line indicates $\rho_c = 0.2050$. Inset shows $\rho_c - \rho(t)$ in semilogarithmic scale.
    (b) Entropy $S_A$ of an interval $A$ as a function of the interval length $|A|$ for several values of $l$. Data from stabilizer simulations of a system of $L = 2048$ qubits, averaged over $> 2000$ circuit realizations. 
    (c) Same data (for $|A|\leq L/2$) in semilogarithmic scale, against fits to $\alpha \log |A| + {\sf const.}$ (dashed lines). 
    (d) Values of the entropy coefficient $\alpha$, extracted from fits to the data in (c), vs AdS radius $l$. We find $\alpha = al+b$ (dashed line). 
    (e) Tripartite mutual information $I_3$ (for consecutive intervals of length $L/4$) vs AdS radius $l$.}
    \label{fig:AdS_states}
\end{figure}

Here we show the results of additional numerical simulations of the states produced from the circuits based on the AdS metric. 
The measurement rate is set to 
\begin{equation}
    \rho(t) = \rho_c \left[ 1 - \left(\frac{l}{l+|t|} \right)^{1/\nu} \right]
\end{equation}
in such a way that $\rho(t=0) = 0$ (the circuit ends when the measurement density vanishes). 
Curves for $\rho(t)$ for different values of $l$ are shown in Fig.~\ref{fig:AdS_states}(a).
We then simulate Clifford circuits with this measurement rate on $L = 2048$ qubits. Note that while in principle one should start the dynamics at $t = -\infty$, in practice we start at $t = -4L$. For $t$ between $-4L$ and $\approx -2L$ the measurement rate is within our error bars on $\rho_c$ [see inset in Fig.~\ref{fig:AdS_states}(a)], so we effectively run a critical circuit of depth $O(L)$ at the start of our dynamics and there is no need to extend the circuit further back.

Results for the entropy of intervals $A$ are shown in Fig.~\ref{fig:AdS_states}(b) for different values of the ``AdS radius'' parameter $l$. The data are found to be in good agreement with the geometric expectation $S_A \simeq \alpha \log |A|$ based on the length of AdS geodesics. 
Moreover, the geometric picture implies a scaling of $\alpha \propto l$ (as the metric is $\diff \ell^2 = l^2 (\diff x^2 + \diff t^2)/t^2$).
We test this conjecture in Fig.~\ref{fig:AdS_states}(d), finding good agreement up to an additive constant, $\alpha = al + b$. The additive constant comes from the background critical dynamics: taking $l\to 0$ yields $\rho(t) = \rho_c$ for all $t\neq 0$, so we recover the universal scaling of entanglement at the MIPT, $S_A \ \frac{c}{3} \log |A|$ where $c$ is the central charge. An offset $b = c/3$ is thus expected. This becomes negligible if we take $l \gg 1$ and look at even larger systems $|A| \gg l$. 

Finally, a signature of the geometric picture for entanglement is the fact that the tripartite mutual information $I_3$ is negative. Plugging in the ansatz $S_A = S_B = S_C = S_{ABC} = \alpha \log(L/4)$, $S_{AB} = S_{BC} = \alpha \log(L/2)$, and $S_{AC} = 2\alpha \log(L/4)$ (the scenario with a connected wedge on $AC$ gives a slightly larger cost $2\alpha\log(L/4) + \alpha\log(3)$) we obtain
\begin{equation}
    I_3(A:B:C) = 2\alpha \log(L/4) - 2\alpha\log(L/2) = -2\log(2)\alpha.
\end{equation}
Thus we expect $I_3 = a'l + b'$ with $a' = -2\ln(2)a$ ($a$ being the fit coefficient in $\alpha = al+b$).
We verify these prediction in Fig.~\ref{fig:AdS_states}(e): $I_3$ is negative, scales approximately linearly with the AdS radius $l$, and the fit coefficients $a$, $a'$ are such that $2\ln(2)a/a' \simeq 1.1$, close to the ideal prediction of 1.


\section{Details of entanglement wedge simulation \label{app:wedge}}

In all of our numerical simulations we use clifford circuits, for which efficient simulation algorithms exist~\cite{aaronson_improved_2004}. Here we explain the specific procedure used, and any tricks which allow for more efficient simulation.

Because we restrict to dual unitary circuits, the gate set is generated by {\textbf{SWAP}} and {\textbf{iSWAP}} two-qubit gates (in a $1:9$ ratio) along with random single-qubit operations applied on their incoming and outgoing legs. The random measurements are performed as single qubit random Pauli measurements. 

The time complexity of performing a 2-qubit gate on a state of $W$ qubits 
(which includes both physical qubits and ancillas) is $O(W)$ while the time complexity of performing a measurement is $O(W^2)$. Because our circuits have a finite density of measurement, it is the latter which dominates the runtime of the simulation. If we run the circuit for a depth $T$ then the total runtime is $O(W^2 \cdot LT)$, where $L$ is the number of physical qubits so that $LT$ is proportional to the total number of measurements. 

To produce a map like Fig.~\ref{fig:entanglement_wedges} we na\"ively need to repeat such a simulation for each point $(x,t)$ of space-time, which would take $O(LT)$ runs of complexity $O(L^3 T)$ each, or a total computation time of $O(L^4T^2)$. For the case of interest, $T \propto L$, this becomes $O(L^6)$. 
This is however not necessary. First of all, we can obtain data for all values of $x$ from a single simulation by simply translating the interval $A$ in the output state (only the relative coordinate matters). Secondly, as we explain next, we can also obtain data for all values of $t$ from a single simulation by making use of ancilla qubits.



The main technical innovation we introduce is making use of the random measurements already present in the dynamics to efficiently insert multiple non-interfering ancillas while simulating the circuit. For each layer, $t$, we prepare an ancilla qubit, $R_t$. 
Out of all qubits that are measured in layer $t$, we pick one at random, $B_t$, and prepare a maximally-entangled Bell pair state $\frac{1}{\sqrt{2}}\ket{00} + \frac{1}{\sqrt{2}}\ket{11}$ on $R_t B_t$. $R_t$ is never acted on for the rest of the dynamics.
At the termination of the circuit we are therefore left with a state on $L + T$ qubits ($L$ system qubits and $T$ ancillas $\{R_t\}$) where the system size is $L$ and the depth of the circuit is $T$.
Simulating this circuit takes $O((L+T)^2 LT) = O(L^4)$. 

To compute the mutual information between the ancilla $R_t$ and a region $A$ on the final timeslice of the circuit we first perform projective measurements in random Pauli bases on all ancillas $R_t'$ with $t' \neq t$. This has the effect of preparing the same final state as if the only ancilla inserted is $R_t$; all of the other ancillas have in effect been prepared in the states they would have been after measurement, at the times they were inserted. From this point we can directly compute the mutual information between $R_t$ and $A$ as desired, without interference from the other inserted ancillas. 


For an arbitrary extensive subsystem each computation of the mutual information takes $O(L^3)$ time, but since our subsystems consist of intervals, or intervals and an ancilla, we can simplify this computation by placing the stabilizers in the clipped gauge~\cite{nahum_quantum_2017}. Let us now assume that $W = L+1$, as we have measured, and traced over all but one ancilla. After placing the stabilizers into the clipped gauge and locating their endpoints, which requires $O(L^3)$ time, the entanglement entropy for an aribtrary interval can be computed in $O(L)$ time.
To compute the mutual information between all intervals of size $l$ and $R_t$ requires computing the entropy of $R_t$, which can be done in $O(L)$ time after preprocessing, the entropy of all size $l$ intervals which can be done in $O(L^2)$ time after preprocessing, and the entropy of all intervals together with $R_t$ which can be done in $O(L^3)$ time including preprocessing when $l = O(L)$ as is the case in our experiments. 

Na\"ively computing the entropy of an interval and the ancilla takes $O(L^3)$ time even with preprocessing because the joint subsystem is not contiguous. To avoid this we insert the ancilla $R_t$ at locations $i \in {\{n \lfloor \frac{2L}{l} \rfloor ~|~ n \in \mathbb{Z} \wedge 0 \leq n \lfloor \frac{L}{l} \rfloor \leq L\}}$ by shifting all qbits from $i$ to $L$ one step to the right and then placing $R_t$ at $i$. After this all intervals of the form $[j, j + l + 1]$ containing $i$ correspond to the original subsystem $[j, j + l] \cup R_t$. Hence for each location $i$ and for all corresponding $j$ such that $i - l \leq j \leq i + l$ we can compute the entropy of $[j, j + l] \cup A$ in $O(L)$ amortized time with $O(L^3)$ preprocessing. Notice that every choice of $j$ is covered by at least one choice of $i$ by design. As there are $\left\lceil\frac{L}{2l}\right\rceil = O(1)$ possible locations for $i$ this corresponds to the aforementioned $O(L^3)$ time. Computing the mutual informations with all $T$ ancillas therefore takes $O(L^4)$ time and is comparable to the time required for state preparation.

This procedure has an overall complexity of $O(L^4)$ whereas the complexity of running a different simulation for each value of $t$ and entangling only one ancilla per run at step $t$ is $O(L^5)$, so we find an $O(L)$ speedup using our procedure. This advantage comes at the expense of some more correlation among different datapoints (as a single circuit realization provides data for all values of $t$).

\end{document}